\documentclass[12pt]{article}
\usepackage[]{graphicx}
\usepackage[]{color}
\usepackage{setspace}
\usepackage{wrapfig}
\graphicspath{{figs/}}

\makeatletter
\def\maxwidth{ %
  \ifdim\Gin@nat@width>\linewidth
    \linewidth
  \else
    \Gin@nat@width
  \fi
}
\makeatother

\definecolor{fgcolor}{rgb}{0.345, 0.345, 0.345}

\usepackage{framed}
\makeatletter
 {\par\unskip\endMakeFramed%
 \at@end@of@kframe}
\makeatother

\definecolor{shadecolor}{rgb}{.97, .97, .97}
\definecolor{messagecolor}{rgb}{0, 0, 0}
\definecolor{warningcolor}{rgb}{1, 0, 1}
\definecolor{errorcolor}{rgb}{1, 0, 0}

\usepackage{alltt}
\usepackage{amsmath}
\usepackage{amsfonts}
\usepackage{natbib}
\usepackage{hyperref}
\hypersetup{
    colorlinks=true,
    linkcolor=blue,
    filecolor=magenta,      
    urlcolor=cyan,
    citecolor=blue,
}

\usepackage{newfloat}
\DeclareFloatingEnvironment[name={Supplementary Figure}]{suppfigure}

\IfFileExists{upquote.sty}{\usepackage{upquote}}{}
\begin{document}
\bibliographystyle{plainnat}
\title{Evolution as a coexistence mechanism:\\ Does genetic architecture matter?}
\author{Sebastian J. Schreiber\footnote{Department of Evolution and Ecology and Center for Population Biology, University of California, Davis USA 95616 e-mail: sschreiber@ucdavis.edu}, Swati Patel\footnote{Department of Evolution and Ecology and Graduate Group in Applied Mathematics, University of California, Davis USA 95616}, and Casey terHorst\footnote{Department of Biology, California State University, Northridge USA 91330}}

\maketitle

\centerline{\large Accepted for publication in \emph{The American Naturalist}}

\begin{abstract}
Species sharing a prey or a predator species may go extinct due to exploitative or apparent competition. We examine whether evolution of the shared species acts as a coexistence mechanism and to what extent the answer depends on the genetic architecture underlying trait evolution. In our models of exploitative and apparent competition, the shared species evolves its defense or prey use. Evolving species are either haploid or diploid. A single locus pleiotropically determines prey nutritional quality and predator attack rates. When pleiotropy is sufficiently antagonistic (e.g. nutritional prey are harder to capture), eco-evolutionary assembly culminates in one of two stable states supporting only two species. When pleiotropy is weakly antagonistic or synergistic,  assembly is intransitive: species-genotype pairs are cyclically displaced by rare invasions of the missing genotypes or species. This intransitivity allows for coexistence if, along its equilibria, the geometric mean of recovery rates exceeds the geometric mean of loss rates of the rare genotypes or species. By affecting these rates, synergistic pleiotropy can mediate coexistence, while antagonistic pleiotropy does not. For diploid populations experiencing weak antagonistic pleiotropy, superadditive allelic contributions to fitness can mitigate coexistence via an eco-evolutionary storage effect. Density-dependence and mutations also promote coexistence. These results highlight how the efficacy of evolution as a coexistence mechanism may depend on the underlying genetic architecture.
\end{abstract}



\newpage

\section*{Introduction}
Evolution has produced an immense diversity of species on earth. When these species share resources or natural enemies, diversity decreases when exploitative competition or apparent competition drives some of them extinct~\citep{grover-97,holt-lawton-93}. For species sharing a common prey or resource (``exploitative competition''), this species loss may be determined by the ``R$^*$ rule'': the species which suppresses the resource to the lower equilibrium density  (R$^*$) excludes the other species~\citep{volterra-28,hsu-hubbell-waltman-77,tilman-82,grover-97,kirk-02,miller-etal-05,wilson-etal-07}. Species sharing a common predator or pathogen may experience apparent competition--an increase of one species' density that leads to an increase in the predator's density and a corresponding decrease in the other prey species' density~\citep{holt-77,holt-lawton-93,holt-etal-94,bonsall-hassell-97,chaneton-bonsall-00,morris-etal-04}. When predators are at sufficiently high densities, the ``$P^*$ rule'' predicts that the prey species supporting the higher equilibrium predator density ($P^*$) excludes the other prey species~\citep{holt-77,holt-lawton-93}.

Ecologists have identified a diversity of mechanisms that can maintain diversity and prevent apparent or exploitative competition from excluding species ~\citep{chesson-00}. Traditionally, these coexistence mechanisms were considered to be of an ecological nature~\citep{chesson-00}, but an increasing number of studies demonstrate that evolutionary changes in traits occur on sufficiently short time scales to influence ecological dynamics~\citep{strauss-etal-08,schoener-11}. 
Several lines of evidence point to the fact that changes in traits may contribute to species coexistence~\citep{lankau-strauss-07,ecology-11b,vasseur-etal-11,amnat-15c}. Plasticity in traits can alter species interactions and increase community stability and coexistence~\citep{vos-etal-04,miner-etal-05}. Inducible defenses in prey species often increase coexistence with predators \citep{vanderstap-etal-08,verschoor-etal-04,petrusek-etal-09}. For example, competition between rotifer species led to the exclusion of the less competitive rotifer, but when a shared predator was present, inducible defenses in the less competitive rotifer led to coexistence of species at both trophic levels~\citep{vanderstap-etal-08}. Similarly, predators who constantly switch strategies to attack the most abundant or the most palatable prey species increase coexistence relative to predators with a fixed behavior~\citep{krivan-03}.

Just as within-generation changes in traits increase stability and coexistence, so do across generation changes due to evolution~\citep{lankau-11}. The evolution of defensive or
predator-avoidance traits in prey can allow species to find enemy free space~\citep{jeffries-lawton-84}. Such trait evolution can increase coexistence between predator and prey~\citep{jones-etal-09,fischer-etal-14,ikegawa-etal-15} and alter the stability of predator-prey cycles \citep{yoshida-etal-03,yoshida-etal-07,becks-etal-12}. For example, populations of intertidal molluscs with strong predator avoidance strategies in response to predatory sunstars showed increased coexistence with predators in natural communities, relative to prey populations with weaker avoidance strategies ~\citep{escobar-navarrete-11}. In systems with intraguild predation \citep{amnat-15c,wang-etal-16} or apparent competition~\citep{ecology-11b,nrm-15}, intraspecific variation or evolution of predator traits can also stabilize communities and lead to species coexistence. 

However, it is not well understood how different genetic architectures of evolving traits may affect the role that eco-evolutionary feedbacks play in facilitating species coexistence~\citep{yamamichi-ellner-16}. Most studies to date have focused on the effects of genetic architecture on purely evolutionary dynamics. In particular, components of the genetic architecture of traits, such as species ploidy, patterns of dominance, pleiotropy, and the distribution of mutational effects, are predicted to affect how species evolve~\citep{hansen-06}.  For example, diploid species tend to have greater genetic variation by virtue of more mutations, but also tend to be less efficient in responding to selection~\citep{otto-gerstein-08}. In addition, theory suggests that traits influenced by multiple loci or multiple alleles make trait dynamics more prone to cycles or even chaos~\citep{seger-88,kopp-gavrilets-06}. While the role of genetic architecture on evolutionary dynamics has been explored, the ecological consequences at the community level due to eco-evolutionary feedbacks remain unexplored. Work on predator-prey co-evolution suggests these consequences may be substantial~\citep{doebeli-97,yamamichi-ellner-16}. For example, predator extinction is more likely when there is dominance at a single diploid locus for a prey trait~\citep{yamamichi-ellner-16} and when the number of loci that contribute to the predator trait is much greater than the number of loci contributing to the prey trait~\citep{doebeli-97}.

Here we explore the role of eco-evolutionary feedbacks and genetic architecture on mediating coexistence for species sharing a common prey or predator species. We fuse classical ecological models of exploitative and apparent competition with classical population genetic models accounting for haploid and diploid genetics, pleiotropy, dominance, and mutation. In these models, pleiotropy occurs ecologically through the simultaneous effects of genes on the attack rate of predators and the nutritional benefit of captured prey. Dominance arises in whether a single copy of an allele is sufficient to defend against a particular predator species or sufficient to effectively attack a particular prey species. We conduct a mathematical analysis that identifies when coexistence of all species and genotypes occurs in the sense of permanence~\citep{hofbauer-sigmund-98}. We also numerically explore to what extent this coexistence occurs via a red queen dynamic, converges to a stable eco-evolutionary state, or is limited by the rate of mutations.

\section*{Models and Methods}
\begin{figure}[t!!]
\begin{center}
\includegraphics[width=4in]{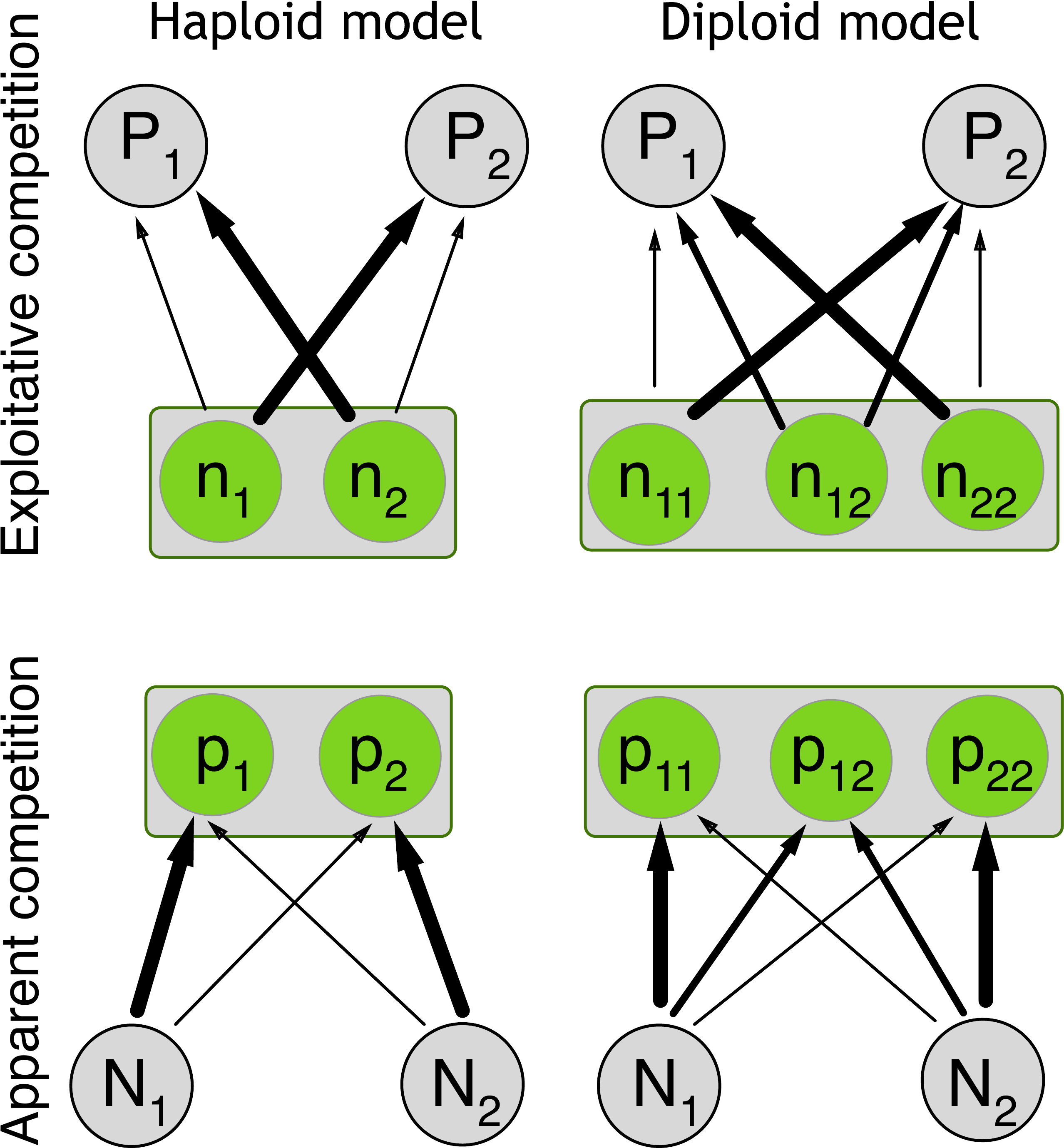}
\end{center}
\caption{Schematics for the exploitative competition (top) and apparent competition (bottom) models. Circles correspond to the evolving genotypes (in green) and the non-evolving species (in gray). Solid black arrows correspond to feeding links with the width of the arrow representing the magnitude of the corresponding per-capita attack rates. }\label{fig:schematic}
\end{figure}

To explore the roles of ecological and genetic structures on eco-evolutionary coexistence mechanisms, we study four models with two ecological and two genetic structures (Fig.~\ref{fig:schematic}). The ecological structures correspond to two classical ecological modules: exploitative and apparent competition. In the exploitative competition module, two predator species (which may be herbivores, predators, parasites, or pathogens) share a common prey (which may be plants, herbivores, or hosts). In the apparent competition module, two prey species share a common predator but do not interact directly. We use Lotka-Volterra equations to model the ecological dynamics in both modules. 

For both ecological modules, the shared species can evolve. Specifically, the shared prey's defense against predation evolves in the exploitative competition module and the shared predator's resource use evolves in the apparent competition module. We model this evolution with both haploid and diploid genetics. For both genetic structures, we assume that only the interspecific interactions drive selection. This assumption allows us to focus on how the countervailing selection pressures from other species, in and of themselves, influence the eco-evolutionary dynamics. Selection occurs at a single locus with two alleles, $A_1,A_2$, where allele $A_i$ provides the best adaptive response to species $i$. For the diploid model, individuals randomly mate and heterozygote individuals are assumed to have intermediate phenotypes. We also investigate how mutation between alleles influences coexistence.

\subsection*{The exploitative competition module}
The exploitative competition model consists of two predatory species with densities $P_1,P_2$ and a common evolving prey. For the haploid version of the model, $n_i$ is the density of prey genotype $A_i$. For the diploid version, $n_{ij}$ is the density of prey genotype $A_iA_j$, and $n_i=2n_{ii}+n_{12}$ is the density of $A_i$ alleles. The total prey density equals $N=n_1+n_2$ for the haploid model and $N=n_{11}+n_{12}+n_{22}=(n_1+n_2)/2$ for the diploid model. Prey individuals live in one of $K$ habitable sites in the landscape  e.g. germination sites, territories, nesting or breeding sites. All prey genotypes produce offspring at a rate $b$ of which a fraction $1-N/K$ survive. All prey individuals die at rate $d$. 

The prey genotype pleiotropically affects both the prey's defense against the predators and its nutritional value. Hence, predator's attack rates and conversion efficiencies, i.e., how much reproductive benefit the predator receives from each prey consumed, depend on the prey genotype: $a_i^\ell$, $c_i^\ell$, respectively, are the attack rate and conversion efficiency of predator $\ell$ on haploid prey genotype $A_i$, and  $a_{ij}^\ell$, $c_{ij}^\ell$, respectively, are the attack rate and conversion efficiency of predator $\ell$ on diploid prey genotype $A_iA_j$. Since conversion efficiencies do not directly affect the prey fitness, selection only directly acts on the defensive trait of the prey. Individuals of predator $\ell$ die at rate $\delta_\ell$. 

Under these assumptions, the haploid dynamics are governed by:

\begin{equation}
\begin{aligned}
\frac{dn_{1}}{dt}&=n_1(b(1-N/K)-d- a_{1}^1 P_1- a_{1}^2 P_2)\\
\frac{dn_{2}}{dt}&=n_2(b(1-N/K)-d- a_{2}^1 P_1- a_{2}^2 P_2)\\
\frac{dP_1}{dt}&=P_1(c_{1}^1 a_{1}^1 n_{1}+c_{2}^1 a_{2}^1 n_{2}-\delta_1)\\
\frac{dP_2}{dt}&=P_2(c_{1}^2 a_{1}^2 n_{1}+c_{2}^2 a_{2}^2 n_{2}-\delta_2)\\
\end{aligned}
\end{equation}

For the diploid model, we define $x_{ij}=n_{ij}/N$ as the frequency of prey genotype $A_iA_j$. If individual prey mate randomly and the prey have a one-to-one sex ratio, then the diploid dynamics satisfy 

\begin{equation}
\begin{aligned}
\frac{dn_{11}}{dt}&=bN((x_{11})^2+x_{11}x_{12}+(x_{12})^2/4)(1-N/K)-dn_{11}- a_{11}^1 n_{11}P_1-a_{11}^2 n_{11}P_2\\
\frac{dn_{22}}{dt}&=bN((x_{22})^2+x_{22}x_{12}+(x_{12})^2/4)(1-N/K)-dn_{22}-a_{22}^1 n_{22}P_1-a_{22}^2 n_{22}P_2\\
\frac{dn_{12}}{dt}&=bN(x_{11}x_{12}+2x_{11}x_{22}+x_{22}x_{12}+(x_{12})^2/2)(1-N/K)-dn_{12}-a_{12}^1 n_{12}P_1-a_{12}^2 n_{12}P_2\\
\frac{dP_1}{dt}&=P_1(c_{11}^1 a_{11}^1 n_{11}+c_{12}^1 a_{12}^1 n_{12}+c_{22}^1 a_{22}^1 n_{22}-\delta_1)\\
\frac{dP_2}{dt}&=P_2(c_{11}^2 a_{11}^2 n_{11}+c_{12}^2 a_{12}^2 n_{12}+c_{22}^2 a_{22}^1 n_{22}-\delta_2)\\
\end{aligned}
\end{equation}

\noindent We also analyze models accounting for mutations of probability $\mu$ for each of the alleles. These modified equations are presented in \ref{AppendixA}. 

\subsection*{Apparent competition model}
The apparent competition model consists of two prey species with densities $N_1, N_2$ and a common evolving predator species. Prey species $i$ exhibits logistic dynamics $\frac{dN_i}{dt}=r_iN_i (1-N_i/K_i)$ in the absence of the predator, where $r_i$ is the intrinsic growth rate and $K_i$ is the carrying capacity. The predator genotypes affect their attack rates and conversion efficiencies with respect to the prey species: $a_\ell^i$, $c_\ell^i$, respectively, are the attack rate and conversion efficiency of the haploid predator $i$ on prey $\ell$, and  $a_\ell^{ij}$, $c^{ij}_\ell$, respectively, are the attack rate and conversion efficiency of the diploid predator genotype $A_iA_j$ on prey $\ell$. Since both affect predator fitness, selection directly acts on both the attack rate and conversion efficiency traits.   Individuals of predator $i$ die at rate $\delta_i$. Equations for the haploid and diploid models are presented in \ref{AppendixB}.

\subsection*{Methods}
Our analyses begin with examining the eco-evolutionary assembly dynamics. That is, we identify which subcommunities of species and genotypes coexist, and how invasions by missing genotypes or species change the ecological or genetic structure of the community. In particular, we analyze subsystems consisting of either three species with only one genotype of the shared species, or two species with all genotypes of the shared species. For each of these subcommunities, we find that the missing species (or allele) can either invade and displace the other species (or allele), or fails to invade. Proofs of these assertions are in the Appendices.

Using the mathematical theory of permanence~\citep{hutson-schmitt-92,hofbauer-sigmund-98,jde-10} in conjunction with our eco-evolutionary assembly analysis, we determine under what conditions all species and genotypes coexist in the sense of permanence. Namely, permanence ensures there is a positive density that all species and genotypes eventually exceed provided all species and genotypes are initially present. This form of coexistence is robust to large rare perturbations as well as frequent small perturbations~\citep{jtb-06}. Our analysis, whose details are presented in the Appendices, explicitly characterizes permanence for all four models with and without mutations. 

To illustrate the main conclusions of our mathematical analysis, we numerically simulate the models with the deSolve package in R \citep{R}. The code for these simulations is available at GitHub~\citep{MBI-code}.

\section*{Results for the exploitative competition module}
\begin{figure}[h!!!]
\begin{center}
\includegraphics[width=0.7\textwidth]{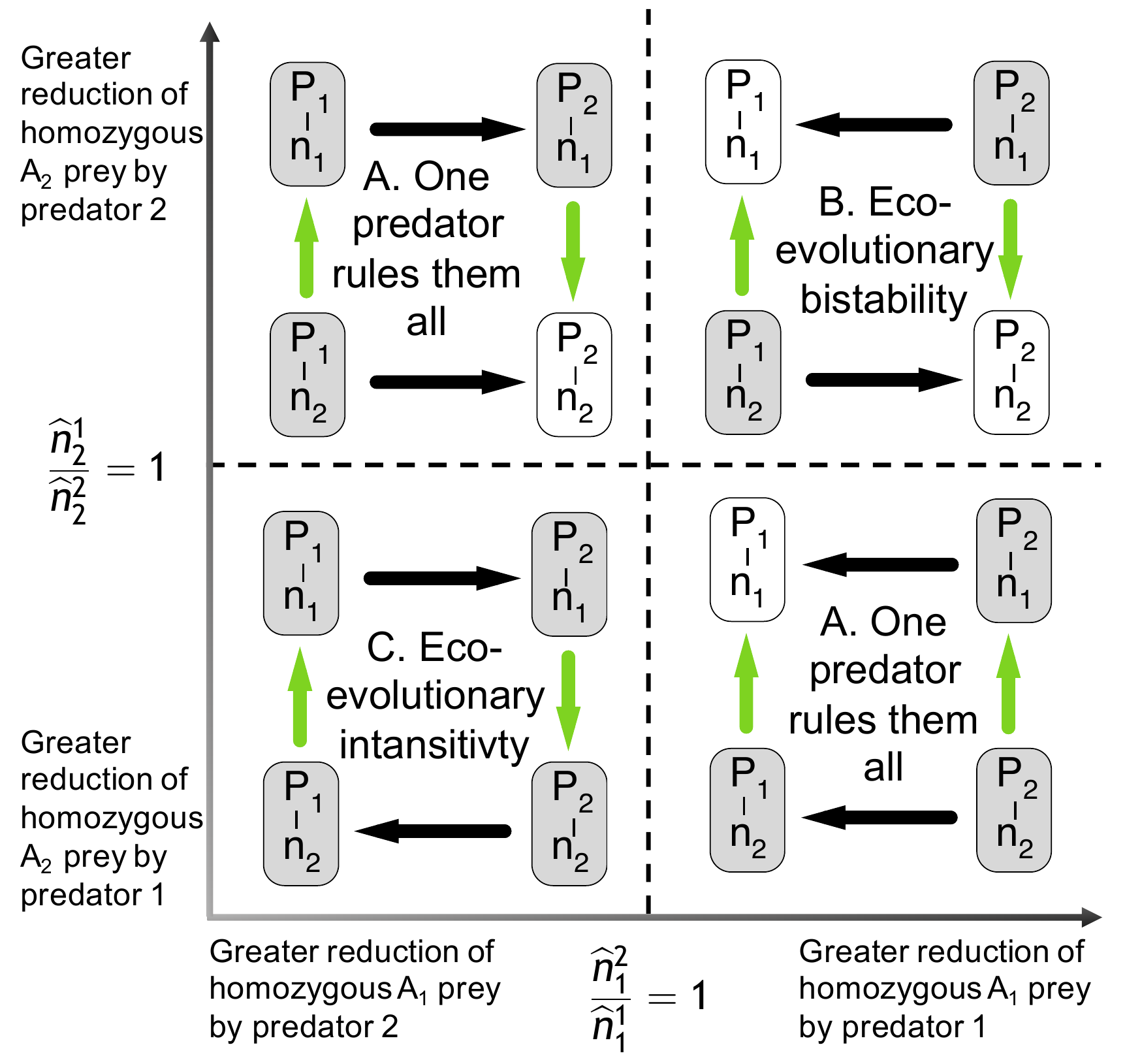}
\end{center}
\caption{Break-even densities determine eco-evolutionary assembly patterns. The horizontal and vertical axes plot the ratio of break even densities for the predators with respect to the homozygous prey genotypes. Whether these ratios are greater than one or less than one determine three types of evolutionary assembly diagrams. In each diagrams, rounded boxes correspond to subcommunities, horizontal black arrows correspond to transitions due to predator invasions, and vertical green arrows correspond to transitions due to invasions of prey alleles.  Non-invadible communities are white boxes, invadible are gray. In A, predator $2$ has lower break even densities with respect to both homozygous prey genotypes. In B, for each homozygous prey genotype, the predator with lower attack rate has the lower break-even density. In C, for each homozygous prey genotype, the predator with higher attack rate has the lower break-even density.}\label{fig:hetero}
\end{figure}

\subsection*{Eco-evolutionary assembly}
Throughout our analysis of the exploitative competition module, we make three assumptions. First, productivity of the system is sufficiently high to ensure that each predator can persist in the presence of each prey genotype. That is, for all $i,j=1,2$, $\widehat n>\widehat n_i^j$ where $\widehat n=K(1-b/d)$ is the prey equilibrium density in the absence of the predators and $\widehat n_i^j$ is the density of the homozygous $A_i$ prey genotype for which predator $j$ has a zero per-capita growth rate. Following the terminology of \citep{hsu-etal-78}, we call $\widehat n_i^j$ the break-even density of predator $j$ with respect to the homozygous $A_i$ prey genotype. In terms of the parameters, these break-even densities equal  $\widehat n_i^j =\frac{\delta_j}{c_i^j a_i^j}$ for the haploid model, and $\widehat n_i^j=\frac{\delta_j}{c_{ii}^j a_{ii}^j} K(1-b/d)> \delta_j$ for the diploid model. Second, we assume that homozygous prey with allele $i$ are defended against predator $i$. That is, $a_{1}^1<a_{2}^1$ and $a_2^2<a_1^2$ for the haploid model and $a_{11}^1<a_{22}^1$ and $a_{22}^2<a_{11}^2$ for the diploid model (widths of arrows in Fig.~\ref{fig:schematic}). Finally, we assume that heterozygous diploid prey exhibit intermediate defense i.e.  $a_{11}^1\le a_{12}^1<a_{22}^1$ and $a_{22}^2\le a_{12}^2<a_{11}^2$. Under these assumptions, if only predator $i$ is in the community, then prey allele $A_i$ always goes to fixation as it provides the best defense against attack by this predator (\ref{AppendixA}). 

When only one prey allele is present, say allele $A_i$, the $R^*$ rule applies~\citep{volterra-28,hsu-etal-78,tilman-82}: the predator with the lower break-even density  $\widehat n_i^j$ with respect to this homozygous prey genotype excludes the other predator (\ref{AppendixA}).  More explicitly, if predator $1$ has the lower break-even density with respect to prey genotype $A_i$ (i.e. $\widehat n_i^1<\widehat n_i^2$), then predator $1$ excludes predator $2$, and if the inequality is reversed, the opposite outcome occurs. 

\begin{figure}[t!!!]
\begin{center}
\includegraphics[width=0.8\textwidth]{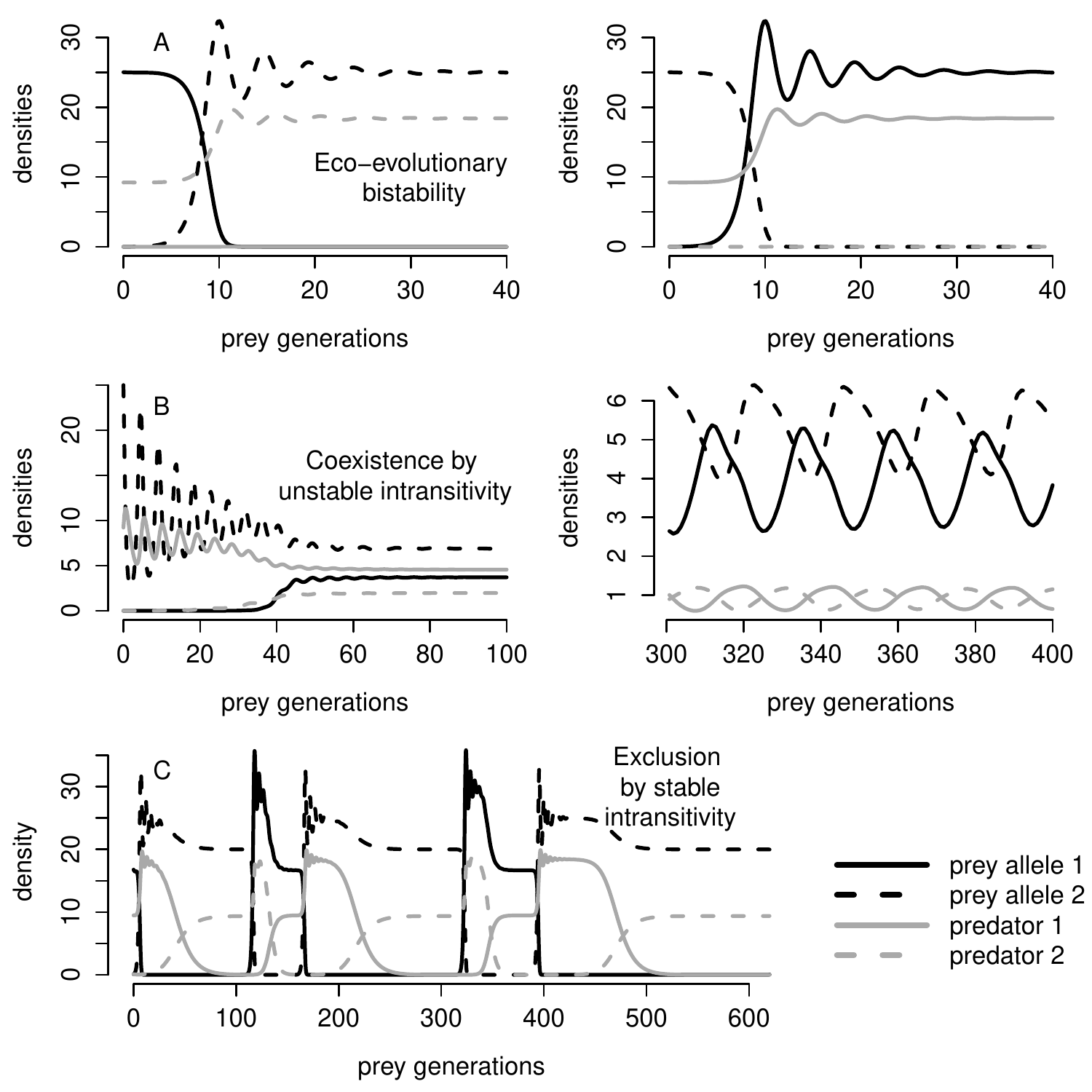}
\end{center}
\caption{Eco-evolutionary outcomes for the diploid exploitative competition module. Predator densities in black, while densities of prey alleles are in gray. In A, an eco-evolutionary bistability--different initial conditions lead to different stable equilibria. In B, two dynamics of an unstable intransitivity: equilibrium and oscillatory coexistence on the left and right, receptively. In C, the dynamics of a stable intransitivity--densities of each species and genotype approach zero in a cyclic fashion.}\label{fig:bistable} 
\end{figure}

The relative values of the break-even prey densities $\widehat n_i^j$ determine three types of eco-evolutionary assembly dynamics (Fig.~\ref{fig:hetero}). First, if one predator has the lower break-even densities with respect to both homozygous prey, then the assembly dynamics culminate in a community consisting of this predator and the associated defended prey genotype (Fig.~\ref{fig:hetero}A). Second, if for each homozygous prey, the predator with the lower attack rate has the lower break-even density (i.e. $\widehat n_1^2>\widehat n_1^1$ and $\widehat n_2^1>\widehat n_2^2$), then the eco-evolutionary feedbacks result in an eco-evolutionary bistability (Figs.~\ref{fig:hetero}B,\ref{fig:bistable}A), in which the stable subcommunities correspond to a predator and the associated defended prey genotype. This outcome only occurs if there is sufficient antagonistic pleiotropy in which the more defended prey genotype for a given predator is more nutritional for that predator. We quantify this pleiotropy using the log ratios of predator conversion efficiencies for undefended to defended prey genotypes. That is, 

\[
\begin{aligned}
\alpha_1&=\log\frac{c_2^1}{c_1^1} \mbox{ and }\alpha_2=\log\frac{c_1^2}{c_2^2} \mbox{ for the haploid model, and }\\
\alpha_1&=\log\frac{c_{22}^1}{c_{11}^1} \mbox{ and }\alpha_2=\log\frac{c_{11}^2}{c_{22}^2} \mbox{ for the diploid model.}
\end{aligned}
\] 

\noindent When $\alpha_i<0$, there is antagonistic pleiotropy as predator $i$ produces more offspring when consuming the defended prey genotype than the undefended prey genotype. When $\alpha_i>0$, there is synergistic pleiotropy as predator $i$ produces fewer offspring when consuming the defended prey genotype. Bistability requires that pleiotropy is sufficiently antagonistic with respect to at least one of the predator species.

Finally, if  for each homozygous prey genotype, the predator with the higher attack rate has the lower break-even density (i.e. $\widehat n_1^2<\widehat n_1^1$ and $\widehat n_2^1<\widehat n_2^2$), then the eco-evolutionary feedbacks result in intransitive assembly dynamics (Fig.~\ref{fig:hetero}C): Predator $2$ can invade the predator $1$-prey allele $A_1$ community and displace predator $1$, then prey allele $A_2$ can invade and fixate, then predator $1$ can invade and displace predator $2$, and finally prey allele $A_1$ can invade and fixate. This outcome occurs for weakly antagonistic to synergistic pleiotropy.

\subsection*{Coexistence}

\begin{figure}[t!!!!]
\begin{center}
\includegraphics[width=0.85\textwidth]{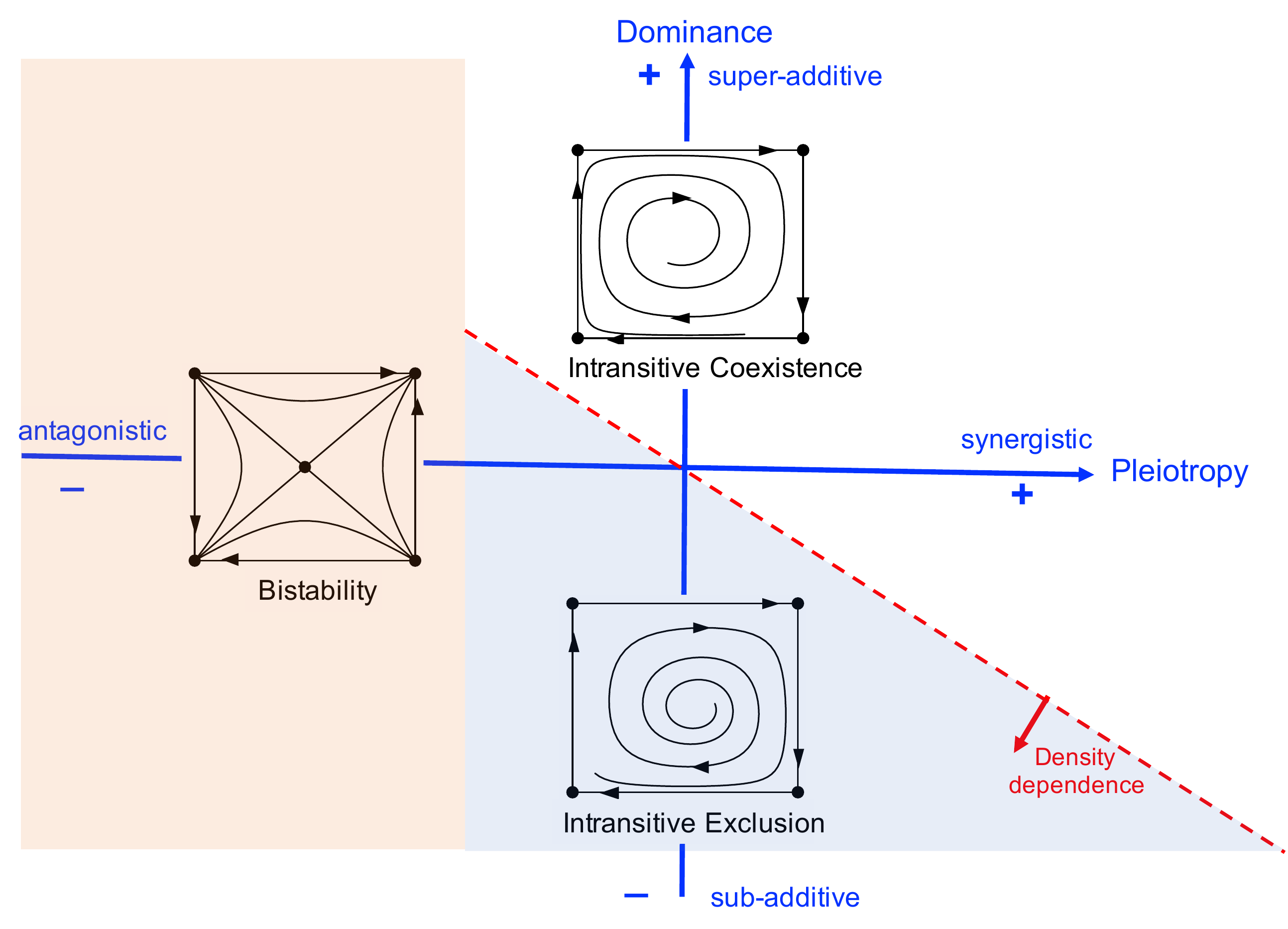}
\end{center}
\caption{The dependency of eco-evolutionary outcomes on the mean pleiotropy and the mean dominance of the defensive alleles in the diploid exploitative competition model. When pleiotropy is sufficiently antagonistic (shaded orange region), the eco-evolutionary dynamics are bistable. When pleiotropy is more synergistic, there is an eco-evolutionary intransitivity. This intransitivity allows for coexistence if the sum of the mean pleiotropy and the mean dominance is positive (white region). Exclusion occurs otherwise (shaded blue region). Density-dependence in the prey increases the region of coexistence (red arrow).}\label{fig:main-bifurcation}
\end{figure}

Coexistence of the predators and the prey genotypes, in the sense of permanence, is only possible for the intransitive assembly dynamics. Whether coexistence occurs depends on whether this intransitivity (a cycle between the four subcommunity equilibria) is unstable or stable. When the intransitivity is unstable (Fig.~\ref{fig:bistable}B), all genotypes and predator species remain bounded away from extinction and may approach an equilibrium (left panel of Fig.~\ref{fig:bistable}B) or exhibit long-term oscillatory behavior (right panel of Fig.~\ref{fig:bistable}B). When the intransitivity is stable, the eco-evolutionary dynamics exhibit increasingly extreme oscillatory dynamics as the community cycles between four eco-evolutionary states dominated by one predator species and one prey allele (Fig.~\ref{fig:bistable}C). From one oscillation to the next, the time spent in each of these states increases, and the frequencies of the rare species and genotypes at each state decrease exponentially fast. Ultimately, for populations of finite size, this leads to the extinction of a predator species and a prey allele.  In \ref{AppendixA}, we derive an explicit condition for coexistence for both models. This condition is summarized graphically in Figure~\ref{fig:main-bifurcation}. Before stating the general form of this condition, we consider the special case of a highly productive system i.e. $K$ is large.

\paragraph{Highly productive systems.}  
For highly productive systems, coexistence for the haploid model occurs if and only if the average pleiotropy is synergistic: 

\begin{equation}\label{eq:hap}
\frac{\alpha_1+\alpha_2}{2}>0.
\end{equation}

\noindent In words, if the defended genotypes are less nutritional on average, then all the species and genotypes coexist at a stable equilibrium or non-equilibrium attractor (Fig.~\ref{fig:bistable}B). Conversely, if the defended genotype is more nutritional on average, then the community is extinction prone: ultimately one predator is excluded and the prey allele least defended to this predator is lost (Fig.~\ref{fig:bistable}C). In Figure~\ref{fig:main-bifurcation}, this condition corresponds to the positive half of the pleiotropy axis.

To state the coexistence condition for the diploid model, we need the following metric of the dominance of the defense alleles: 

\[
\beta_1 = \log\left(\frac{a_{22}^1 - a_{12}^1}{a_{12}^1 - a_{11}^1}\right)
\mbox{ and }\beta_2 = \log\left(\frac{a_{11}^2 - a_{12}^2}{a_{12}^2 - a_{22}^2}\right).
\] 

If $\beta_i=-\infty$, then predator $i$'s attack rate on the heterozygote is the same as its attack rate on the undefended genotype ($a_{12}^i=a_{jj}^i$ with $j\neq i$). Hence, in this case, the defensive allele $i$ is recessive: only individuals with both copies of the defensive allele are defended against predator $i$. Alternatively, if $\beta_i=\infty$, then allele $i$ is dominant: one copy ensures defense against predator $i$. If $\beta_1>0$ and $\beta_2>0$, then the more beneficial allele is dominant with respect to defense against each predator~\citep{rose-82,curtsinger-etal-94}. In particular, if $\beta_1=\beta_2=\infty$, then the alleles are co-dominant: heterozygotes are fully def{}ended against both predator species.  If $\beta_i=0$, then the alleles contribute additively to defense against predator $i$ i.e. the attack rate $a_{12}^i$ on heterozygotes is at the midpoint $(a_{11}^i+a_{22}^i)/2$ of attack rates of the homozygotes.

Coexistence for the diploid model occurs if and only if 

\begin{equation}\label{eq:dip}
\frac{\alpha_1+\alpha_2}{2}+\frac{\beta_1+\beta_2}{2}>0. 
\end{equation}

In words, the sum of the mean pleiotropy and the mean dominance must be positive for coexistence to occur (unshaded region in Fig.~\ref{fig:main-bifurcation}). In the special case that the allelic contributions are additive (i.e. $\beta_1=\beta_2=0$), the diploid coexistence criterion \eqref{eq:dip} reduces to the haploid criterion \eqref{eq:hap}. When the allelic contributions are non-additive, diploidy  can either facilitate or inhibit coexistence. Facilitation is greatest when the alleles are co-dominant with respect to predator defense. Inhibition is greatest when both alleles are recessive with respect to predator defense. 

\paragraph{The general condition for all levels of productivity.} At lower productivity levels, the coexistence condition involves density-dependent ``correction factors.'' We present these correction factors and the general coexistence condition for the diploid model; the haploid coexistence condition corresponds to the coexistence condition for diploids with additive genetics. The density-dependent correction factors are given by 

\[
\gamma_1 =\log \frac{a_{22}^1 \widehat p_{22}^1}{a_{11}^1 \widehat p_{11}^1}= \log \frac{b(1-\widehat n_{22}^1/K)-d}{b(1- \widehat n_{11}^1/K)-d}\mbox{ and }\gamma_2=\log\frac{a_{11}^2 \widehat p_{11}^2}{a_{22}^2 \widehat p_{22}^2} =\log \frac{b(1- \widehat n_{11}^2/K)-d}{b(1- \widehat n_{22}^2/K)-d}.
\]

Namely, for the subsystems with predator $i$, $\gamma_i$ is the log ratio of the equilibrium predation rate on a population of undefended prey to the equilibrium predation rate on a population of defended prey. The equivalence between $b(1-\widehat n_{ii}^j/K)-d$ and $a_{ii}^j\widehat p_{ii}^j$ follows from the prey per-capita growth rates equaling zero at equilibria. As coexistence is only possible with the intransitive eco-evolutionary assembly dynamics (i.e. $\widehat n_{11}^1> \widehat n_{11}^2$ and $\widehat n_{22}^2> \widehat n_{22}^1$), these density-dependent correction factors $\gamma_i$ are always positive. Furthermore, the correction factors $\gamma_i$ are decreasing functions of $K$ and in the limit of high productivity approach a value of zero. 

The general coexistence criterion for the diploid model with the density-dependent correction factors $\gamma_i$ is 
\begin{equation}\label{eq:general}
\frac{\alpha_1+\alpha_2}{2}+\frac{\beta_1+\beta_2}{2}+\frac{\gamma_1+\gamma_2}{2}>0.
\end{equation}
Hence, density-dependence always makes coexistence more likely (red lines in Fig.~\ref{fig:main-bifurcation}). An important special case occurs when  the predator conversion efficiencies are equal among all prey genotypes (i.e. $\alpha_1=\alpha_2=0$). When this occurs and the prey genetics are additive (i.e. $\beta_1=\beta_2=0$), the coexistence criterion is always satisfied due to the density-dependent correction factor. 

\begin{figure}[t!!!]
\begin{center}
\includegraphics[width=0.75\textwidth]{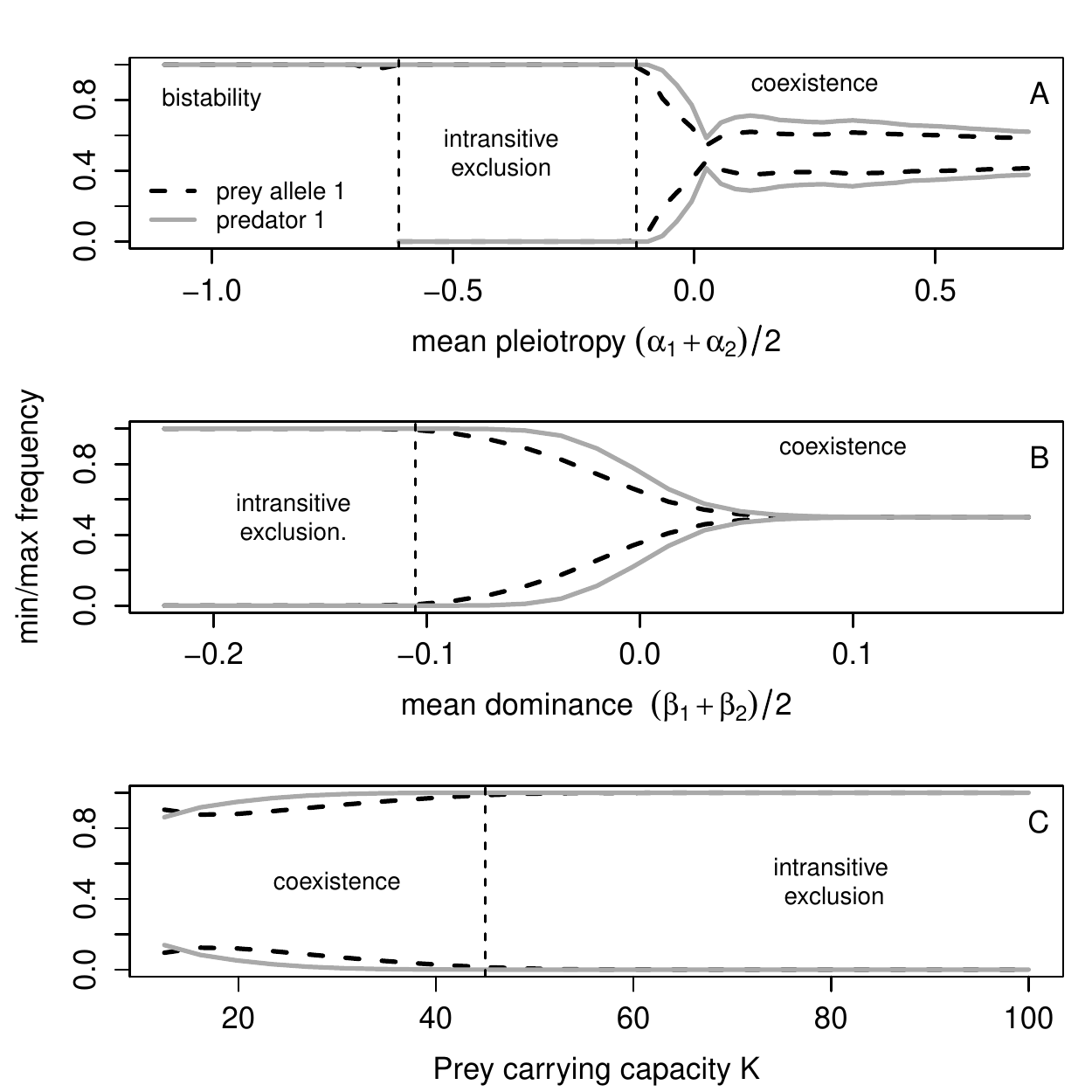}
\end{center}
\caption{Long-term minimum and maximum frequencies as a function of the mean pleiotropy (A), the mean dominance (B) and the prey carrying capacity (C). In A and C, the genetics are additive. In B and C, there is no pleiotropy.  In C, both predator species fail to persist when $K\le 12.5$. Parameter values: $b=1$, $d=0.2$, $\delta_1=\delta_2=0.1$, $c_{11}^2=c_{22}^1=0.2$, $a_{11}^2=a_{22}^1=0.08$, and $a_{11}^1=a_{22}^2=0.04$.  In A, $c_{11}^1=c_{22}^2$ vary between $0$ and $0.6$, $K=400$, $a_{12}^i=(a_{11}^i+a_{22}^i)/2$. In B, $c_{ij}^\ell=0.2$ for all $i,j,\ell$ and $K=400$. In C, $c_{ij}^\ell=0.2$ for all $i,j,\ell$ and $a_{12}^i=(a_{11}^i+a_{22}^i)/2$.}\label{fig:bifurcations}
\end{figure}

Figure~\ref{fig:bifurcations} illustrates several of our analytical results numerically. As predicted by our analysis for additive genetics, coexistence occurs for synergistic pleiotropy and mildly antagonistic pleiotropy due to prey-density dependence (Fig.~\ref{fig:bifurcations}A).  Exclusion through a stable intransitivity occurs with intermediate antagonistic pleiotropy, and exclusion through a bistability occurs with strong antagonistic pleiotropy. Alternatively, even if there is antagonistic pleiotropy, coexistence occurs if there is, on average, sufficiently strong dominance in the defensive alleles, and intransitive exclusion occurs otherwise (Fig.~\ref{fig:bifurcations}B). Finally, prey density-dependence (low $K$) can promote coexistence provided  the prey carrying capacity $K$ is sufficiently high to support both predators (Fig.~\ref{fig:bifurcations}C).

\paragraph{Mutation limited coexistence} When there is a positive mutation probability $\mu>0$ and an eco-evolutionary intransitivity, the species always coexist (\ref{AppendixA}). However, when exclusion occurs without mutation, the coexistence is mutation limited in the sense that the populations exhibit oscillations where the minimal densities of each species and genotype are on the order of the mutation probability (Supplementary Fig.~\ref{fig:haploid-mutation}). 

\section*{Results for the apparent competition module}

Our analysis of the apparent competition module makes three assumptions. First, productivity of the system is sufficiently high (i.e. $K_i\gg 1$ for $i=1,2$) to ensure that the $P^*$ rule holds~\citep{holt-lawton-93}. Without this assumption, the prey species can coexist as predator densities remain too low to cause exclusion. Second, we assume that predator allele $i$ is most adapted to exploiting prey $i$. That is, $c_1^1a_{1}^1>c_1^2a_{1}^2$ and $c_2^2a_2^2>c_2^1a_2^1$ for the haploid model and $c^{11}_1a^{11}_1>c^{22}_1a^{22}_1$ and $c_2^{22}a^{22}_2>c_2^{11}a^{11}_2$ for the diploid model. Finally, for the diploid model, the heterozygous individuals have intermediate phenotypes i.e.  $c_1^{11} a^{11}_1\ge c_1^{12} a^{12}_1>c_1^{22}a_1^{22}$ and $c_2^{22}a_2^{22}\ge c_2^{12}a^{12}_2>c_2^{11}a_2^{11}$. The analysis for this module is presented in (\ref{AppendixB}).

Under these assumptions, if only prey species $i$ is present, then predator allele $A_i$ goes to fixation as it has the lower break-even density with respect to prey $i$ ($\widehat N_{i}^{ii}=\delta/(c_i^{ii}a_i^{ii})$ for the diploid model and  $\widehat N_{i}^{i}=\delta/(c_i^{i}a_i^{i})$ for the haploid model). Alternatively, when only one predator allele is present, say allele $A_i$, the prey which can support a larger equilibrium density of this predator genotype excludes the other prey species. As in the exploitative competition model, there are three types of eco-evolutionary assembly diagrams (Supplementary Fig.~\ref{fig:assemble-apparent}). First, if one prey species supports higher equilibrium densities of both homozygous predator genotypes than the other prey species, then eco-evolutionary assembly always culminates in a community consisting of this prey species and the predator allele specialized on this prey species (Fig.~\ref{fig:assemble-apparent}A). Second, if each prey species supports a higher equilibrium density of the predator genotype least adapted to it, then the assembly dynamics exhibit a bistability (Fig.~\ref{fig:assemble-apparent}B). This outcome is only possible if there is some antagonistic pleiotropy in the sense that a predator genotype adapted to capturing one prey species receives more nutritional reward for capturing individuals of the other  species. As in the exploitative competition model, we quantify this pleiotropy with the log ratio of the conversion efficiency of the least adapted predator genotype to the most adapted predator genotype with respect to prey $i$:

\[
\alpha_1 = \log \frac{c_1^{2}}{c^{1}_1} \mbox{ and } \alpha_2 = \log\frac{c_2^{1}}{c^{2}_2} \mbox{ for haploids, and } \alpha_1 = \log\frac{c_1^{22}}{c^{11}_1}  \mbox{ and }  \alpha_2 = \log\frac{c_2^{11}}{c^{22}_2} \mbox{ for diploids}.
\]

\noindent Synergistic pleiotropy occurs when $\alpha_i>0$. Finally, the eco-evolutionary assembly dynamics are intransitive when each prey species supports a higher equilibrium density of the predator genotype most adapted to it (Fig.~\ref{fig:assemble-apparent}C). This outcome occurs when pleiotropy is weakly antagonistic or synergistic.

Coexistence is only possible in the case of intransitive eco-evolutionary assembly dynamics, and, for the diploid model, depends on the dominance of allele $A_i$:

\[
\beta_i =\log \frac{|c_i^{jj}a_i^{jj}-c_i^{12}a_i^{12}|}{|c_i^{ii}a_i^{ii}-c_i^{12}a_i^{12}|}.
\]

\noindent As before, $\beta_i=-\infty,0,\infty$ corresponds to when the $A_i$ allele is recessive, additive, and dominant, respectively. For the diploid model, coexistence only occurs if 

\begin{equation}\label{eq:apparent}
\frac{\alpha_1+\alpha_2}{2}+\frac{\beta_1+\beta_2}{2}>0.
\end{equation}

\noindent In words, the sum of the mean pleiotropy and the mean dominance of the predator alleles is positive. For the haploid model, the coexistence condition coincides with additive genetics case for the diploid model: $(\alpha_1+\alpha_2)/2>0$.

\section*{Discussion}
Empirical studies and theory have demonstrated that species sharing a prey species or a predator species may be driven to extinction due to the negative indirect effects of exploitative or apparent competition~\citep{volterra-28,macarthur-72,holt-77,hsu-hubbell-waltman-77,hsu-etal-78,tilman-82,holt-lawton-93,bonsall-hassell-97,wilson-etal-07}. Our analysis demonstrates that evolution of the shared species can mitigate these negative indirect effects, and, as a consequence, simultaneously facilitate species coexistence and maintain genetic polymorphisms. This requires trade-offs in the ability to defend against multiple predators, or the ability to attack multiple prey. While such trade-offs are common in nature~\citep{schluter-grant-84,norton-91, sih-etal-98,svanback-eklov-03,bolnick-smith-04}, whether or not they lead to eco-evolutionary feedbacks promoting diversity depends critically on the genetic architecture underlying these trade-offs. Our analysis reveals that pleiotropy, ploidy, dominance, and mutation rates influence how the communities assemble, whether all species and genotypes coexist, and whether this coexistence occurs at a stable equilibrium. 

\subsubsection*{Synergistic pleiotropy promotes coexistence.}
Ecological pleiotropy occurs when a single trait or gene influences multiple components of the ecological dynamics~\citep{strauss-irwin-04,delong-17}. In our models, this pleiotropy arises from the simultaneous effects of alleles on the rates at which different predators capture individual prey and the nutritional quality of the captured prey to different predators. For example, body size can simultaneously affect both capture rate and nutritional quality. When larger prey provide more nutrition and are easier to capture (e.g. increased visibility of prey~\citep{brooks-dodson-65}), the ecological pleiotropy is synergistic~\citep{mcgee-etal-16}. Smaller body sizes make evolving prey harder to capture and may decrease predator number due to providing lower nutritional rewards. In contrast, when larger prey provide greater nutritional rewards, but are more difficult to capture (e.g. by gape-limited fish, birds, or zooplankton predators~\citep{gliwicz-umana-94,persson-etal-96,janzen-etal-00}), the ecological pleiotropy is antagonistic~\citep{paaby-rockman-13}. Larger body size of an evolving prey lowers attack rates but may increase predator numbers by providing greater nutritional rewards.

Ecological pleiotropy determines how the community structure changes due to rare introductions of missing species or genotypes. If the ecological pleiotropy is sufficiently antagonistic ($\alpha_i<0$), the eco-evolutionary assembly dynamics culminate in one of two stable states, depending on the initial community state. Each stable state consists of one non-evolving species and the genotype of the evolving species best adapted to interacting with this species. For example, in the exploitative competition module, antagonistic pleiotropy allows the predator to suppress defended prey genotypes to a lower equilibrium abundance than the other predator species, and this prevents the invasion of the other predator courtesy of the $R^*$ rule~\citep{tilman-82}. In contrast, if the ecological pleiotropy is synergistic ($\alpha_i>0$), non-existent, or weakly antagonistic, then the eco-evolutionary assembly is intransitive: species-genotype pairs get successively displaced in a cyclic fashion due to rare invasions of the missing genotypes or species. Intransitivities in assembly have been observed in ecological models of interacting competitors~\citep{may-leonard-75,yodzis-78,allesina-levine-11} and coevolutionary models of victim-exploiter interactions~\citep{seger-88, gavrilets-98, kopp-gavrilets-06}. Unlike these models, the intransitivities in our models stem from an interdigitation of ecological displacements and selective sweeps of more adapted genotypes. This type of interdigitation has been observed in the empirical work of \citet{lankau-strauss-07} on genotypes of \emph{Brassica nigra} that produce different levels of sinigrin (an allelochemical that kills mycorrhizal fungi that are beneficial to other plant species). \emph{B. nigra} genotypes with high sinigrin concentrations are able to invade diverse communities of other plant species, but patches of high sinigrin \emph{B. nigra} can be invaded by low sinigrin \emph{B. nigra} genotypes that grow quickly because they do not invest energy in costly sinigrin production. These low sinigrin patches are easily invaded by other plant species, resulting in the same type of eco-evo intransitivity that we observe in our model.    

When the eco-evolutionary assembly dynamics are intransitive, coexistence or the loss of multiple species and genotypes may occur. Which outcome occurs depends on the per-capita growth rates of the species and genotypes when they are rare in the community. Unlike classical coexistence theory~\citep{chesson-00}, positive per-capita growth rates of each species or genotype when rare (mutual invasibility) isn't required for coexistence. Indeed, the intransitive assembly in our models is governed by equilibria supporting one genotype and one non-evolving species. At these equilibria, one of the missing species or genotypes has a negative per-capita growth rate while the other has a positive per-capita growth rate. The absolute value of the negative per-capita growth rate (the loss rate) determines how quickly one rare genotype or species decreases, while the positive per-capita growth rate (the recovery rate) determines how quickly the other rare species or genotype increases. Coexistence requires that the geometric mean of the recovery rates is greater than the geometric mean of the loss rates for the equilibria along this intransitivity (\ref{AppendixA}, \ref{AppendixB}, \citet{hofbauer-sigmund-98}). Why geometric means? Heuristically, in the initial phase of community establishment, each species and genotype experiences a fluctuating environment as the composition of the community changes from one equilibrium to the next. The geometric means capture the average rate at which species or genotypes increase or are lost as the community composition fluctuates. Coexistence occurs when the recovery rates dominate over the loss rates. Hence, coexistence is promoted by mechanisms that either increase recovery rates or decrease loss rates. 

Synergistic pleiotropy simultaneously increases the recovery rates and decreases the loss rates and hence, can promote coexistence. For example, in the exploitative competition module, when the community is dominated by one predator and the defended prey genotype, synergistic pleiotropy reduces the density of this predator and, thereby, decreases selection against the undefended genotype. Synergistic pleiotropy also limits the predator's ability to suppress the density of this genotype and, thereby, increases the recovery rate of the other predator. Antagonistic pleiotropy has the opposite effects and, consequently, tends to disrupt coexistence. Density-dependence can mitigate the effects of antagonistic pleiotropy. For example, in the exploitative competition module, density-dependence in the evolving prey simultaneously decreases loss rates and increases recovery rates. This mitigation allows for an eco-evolutionary counterpart to the paradox of enrichment~\citep{rosenzweig-71}: increasing the carrying capacity of the prey destabilizes the intransitivity and one species and one genotype are lost (Fig.~\ref{fig:bifurcations}C).

\subsubsection*{Comparisons to earlier ecological theory}

In the case of the haploid model, these results parallel findings from earlier ecological studies of consumer species competing for resources~\citep{leon-tumpson-75,tilman-80,leibold-96,oikos-04}. In all of these earlier studies, the same necessary condition for coexistence was found. Namely, each predator species has a lower break-even density than the other predator with respect to one of the prey species. When this occurs, there is a coexistence equilibrium at which \citep[p.194]{leon-tumpson-75} ``each species is limited only by one resource which is different from that limiting the other species.'' This condition isn't sufficient for coexistence, however, when the resources can be driven extinct via apparent competition. Instead, this classical coexistence condition is only sufficient to ensure there is an intransitivity in the assembly dynamics. \citet{oikos-04} studied these intransitive dynamics for ecological models of two competing prey species which are exploited by two predator species. They found that for highly productive systems (i.e. weak density-dependence in the prey), ``coexistence required the predators convert their preferred prey at least as efficiently their less preferred prey'' i.e. synergistic pleiotropy if one views the two prey species as two prey genotypes. Our results extend this result by showing that density-dependence always has a positive effect on coexistence. Hence, density-dependence can mitigate coexistence even if the predators convert their preferred prey less efficiently then their less preferred prey.

\subsubsection*{Dominance of defensive alleles promotes coexistence}

Our analysis highlights that the inefficacy of selection for diploid populations compared to haploid populations can stabilize communities through a genetic ``storage effect''. As rare alleles are masked in heterozygotes of diploid populations~\citep{otto-gerstein-08,gerstein-otto-09}, recovery rates (respectively, loss rates) of rare alleles are lower (respectively, higher) in diploid populations than in haploid populations. The net effect of this selective inefficacy on coexistence depends on the dominance of the alleles, i.e., whether allelic contributions to defense or resource-use are superadditive or subadditive. For the exploitative competition module, superadditivity in prey fitness occurs when the per-capita attack rate of a predator on heterozygous genotypes is lower than the average per-capita attack rate on homozygous genotypes. That is, the heterozygotes are better defended than average homozygous genotype. Superadditivity with respect to both predators is caused by a beneficial reversal of dominance: when a single locus contributes to two aspects of fitness (in this case, defense against two different predators), the more advantageous allele is dominant~\citep{rose-82,curtsinger-etal-94}. Superadditivity ensures the stabilizing effects of reduced loss rates outweigh the destabilizing effects of reduced recovery rates. These reduced loss rates help store alleles during periods in which they do not provide a fitness benefit such as defense against a particular predator--an eco-evolutionary analog of the storage effect~\citep{chesson-warner-81,chesson-94}. A related stabilizing mechanism exists for maintaining genetic polymorphisms in fluctuating environments~\citep{gillespie-langley-74,gillespie-78}. \citet{gillespie-78}'s SAS-CFF model demonstrates that superadditivity for heterozygote fitness (the concave fitness function--CFF) in a stochastic environment increases the geometric mean of fitness (the stochastic additive scale--SAS) of heterozygotes via Jensen's inequality~\citep{jensen-1906,ruel-ayres-99} and, thereby, heterozygotes persist and allelic diversity is maintained. 

When synergistic pleiotropy or superadditivity of heterozygotes is sufficiently strong, our simulations suggest that species coexistence occurs at a stable equilibrium. At this equilibrium, eco-evolutionary feedbacks minimize fitness differences among the non-evolving species. For example, in the exploitative competition module, the prey genotypic frequencies at the coexistence equilibrium are such that both predator species, in isolation, have equal break-even densities~\citep{hsu-etal-78}. Thus, eco-evolutionary feedbacks equalize the fitness differences of the competing predators~\citep{chesson-00,lankau-11}. This equalization, in and of itself, only allows for neutral coexistence in which small levels of demographic or environmental stochasticity can result in species loss~\citep{chesson-88,adler-etal-07}. Evolution, however, stabilizes coexistence by favoring whichever species becomes less common~\citep{lankau-11}. While these results highlight an important ecological feature of the coexistence equilibrium, we haven't studied the genetic features of this equilibrium. However, we can gain some insights from \citet{wilson-turelli-86} who studied the evolution of resource use for a diploid consumer population with two implicitly defined resources. As in our model, resource use is determined by two alleles at a single locus and there is a trade-off between using one resource and using the other resource. They found that superadditive contributions of alleles to resource-uptake lead to a stable polymorphic equilibrium at which heterozygotes are the most fit (overdominance), while subadditive contributions can result in a stable polymorphic equilibrium with heterozygote disadvantage (underdominance). As their model doesn't explicitly account for resource dynamics or ecological pleiotropy, it remains to be seen if their conclusions extend to our model and what role, if any, ecological pleiotropy plays in determining the relative fitness of heterozygotes at polymorphic equilibria.   

\subsubsection*{Mutational rescue}

When antagonistic pleiotropy or subadditivity of heterozygotes are barriers to coexistence, mutations can serve as a stabilizing mechanism by rescuing alleles that otherwise would be lost. However, this rescue effect only permits a fragile form of oscillatory coexistence. That is, our simulations suggest that the densities of the non-evolving species and the frequencies of genotypes repeatedly reach levels proportional to the mutation rates and, consequently, may lead to permanent loss of the non-evolving species via demographic stochasticity. This ``mutation limited'' form of coexistence was observed in two-species coevolutionary models  of host-parasites \citep{seger-88} and mimicry \citep{gavrilets-98}. In both of these earlier studies, both species were evolving at a single diallelic locus and the authors numerically showed that in cases where one allele in each species was lost with no mutation, low mutation enabled cycles in allele frequencies close to fixation of alternating alleles. Our work provides an analytic demonstration of mutation as a coexistence mechanism.

\subsubsection*{Future challenges and opportunities}

Our results highlight several opportunities for empirical and theoretical work. Although there is considerable work examining how a single trait affects multiple components of species interactions, much of it has not invoked the term “ecological pleiotropy” (reviewed in ~\citet{strauss-irwin-04}). Thus, it is unclear whether ecological pleiotropy is synergistic or antagonistic in previous work. When ecological pleiotropy is sufficiently antagonistic, it creates an eco-evolutionary mismatch, in which the predator against which prey are most defended is also the predator that suppresses them to the lowest density. This eco-evolutionary mismatch disrupts the opportunity for eco-evolutionary feedbacks to facilitate coexistence. Knowing the prevalence of synergistic versus antagonistic ecological pleiotropy in natural communities would provide greater knowledge of the extent to which eco-evo feedbacks affect species diversity. Additionally, we have shown super-additive selection on traits by multiple species can lead to a genetic storage effect. However few studies have measured non-additive selection, although data likely exists to do so in many different systems~\citep{terhorst-etal-15}. Finally, our models assumed the evolutionary dynamics are governed by two alleles at a single locus. However, multiple loci likely determine the traits that govern interactions with multiple species; accounting for multiple loci has the potential to change eco-evolutionary dynamics by altering the capacity for species to respond to selection pressures \citep{seger-88, doebeli-97, kopp-gavrilets-06}. Understanding how this additional genetic complexity, which allows for recombination and epistasis, influences our conclusions remains to be tested. 

\subsubsection*{Concluding remarks}

Our results show that eco-evolutionary feedbacks can act as a coexistence mechanism and that the strength of this mechanism depends on underlying genetics. Synergistic pleiotropy, density-dependence, diploidy with dominance of the better adapted allele, and mutation can act as stabilizing mechanisms. Stabilization occurs either by increasing the rate at which rare genotypes or species recover, or by slowing the rate at which rare species or genotypes are lost and, thereby, allow sufficient time for other eco-evolutionary feedbacks to rescue these at-risk species or genotypes. The extent to which these genetic details influence the stability of natural communities, which are inherently more complex ecologically and genetically, remains to be seen.

\paragraph{Acknowledgments.} The authors thank Michael Turelli for suggesting the relevance of the storage effect and introducing us to the SAS-CCF models, and the Mathematical Biosciences Institute for hosting a workshop on eco-evolutionary feedbacks whose participants provided encouragement in the initial phase of this work. Kelsey Lyberger and Sam Fleisher  Two anonymous reviewers and the editor provided extensive constructive comments on the manuscript which greatly improved the presentation. This research was funded by U.S. National Science Foundation Grants DMS-1313418, DMS-1312490 to SJS and CT. 

\bibliography{ecoevo}

\newpage
\renewcommand{\thesection}{Appendix S\arabic{section}}
\section{Analysis of exploitative competition models}\label{AppendixA}
In this Appendix, we determine the conditions necessary for coexistence, in the sense of permanence. The first step of the analysis is to first study the dynamics of subsystems corresponding to a single prey allele with both predators and all prey alleles with single predator. Under the assumptions stated in the main text, this analysis reveals that there is a heteroclinic cycle connecting four equilibria of the boundary of the state space i.e. $\mathbb{R}^4_+=[0,\infty)^4$ for the haploid model and $\mathbb{R}^5_+=[0,\infty)^5$ for the diploid model. The second step of the analysis determines the conditions under which this heteroclinic cycle is repelling (in which case coexistence occurs) or attracting (in which case the system is extinction prone). The first half of the first step of the analysis (i.e. studying the subsystems with a single prey allele with one or both predators) can be carried in parallel for both models. The remainder of the analysis is model specific.

Two common forms of subsystems of the haploid and diploid models is a homozygous prey with one or two predators. As the analysis of these subsystems are identical (only need to replace $i$ with $ii$ for the diploid models), we focus on the haploid case. For the subsystem consisting of a single prey allele, say $i$, and single predator, say $\ell$, the haploid model reduces to the classical Lotka-Volterra predator-prey model

\[
\begin{aligned}
\frac{dn_i}{dt}&= n_i b (1- n_i/K)-d n_i - a_{i}^\ell n_i P_\ell\\
\frac{dP_\ell}{dt}& = c_{i}^\ell a_{i}^\ell n_i P_\ell - \delta_\ell P_\ell.
\end{aligned}
\]

\noindent The prey and predator coexist if and only if $c_i^\ell a_i^\ell K>\delta_\ell$ which we assume holds true throughout our analysis. Coexistence occurs around a globally stable equilibrium given by

\[
\widehat n_i^\ell =\frac{\delta_\ell}{c_i^\ell a_i^\ell} \mbox{ and }\widehat P_i^\ell = \frac{b}{a_i^\ell}(1-\widehat n_i^\ell/K)-\frac{d}{a_i^\ell}.
\]

\noindent For the subsystem consisting of a single prey allele, say $i$, and both predators, the models reduce to a three species Lotka-Volterra model (haploid model shown): 

\[
\begin{aligned}
\frac{dn_i}{dt}&= n_i b (1- n_i/K)-d n_i - a_{i}^1 n_i P_1- a_{i}^2 n_i P_2\\
\frac{dP_1}{dt}& = c_{i}^1 a_{i}^1 n_i P_1 - \delta_1 P_1\\
\frac{dP_2}{dt}& = c_{i}^2 a_{i}^2 n_i P_2 - \delta_2 P_2\\
\end{aligned}
\]

\noindent A classical argument due to \citep{volterra-28} implies that the predator species that can reduce the prey to the lower equilibrium density excludes the other species (see, e.g., Section 5.4 of \citet{hofbauer-sigmund-98}). The remainder of the analysis is carried separately for the haploid and diploid models. 

\subsection*{The haploid case}
The haploid model is a Lotka-Volterra model and, consequently, we can use the basic results about these models described by \citep{hofbauer-sigmund-98}.  We continue by examining the dynamics of both prey genotypes with a single predator, say predator species $1$. For this subsystem, our assumption that $a_1^1< a_2^1$ implies there is no coexistence equilibrium. Hence, \citep[Theorem 5.2.1]{hofbauer-sigmund-98} implies that all trajectories starting with all species converge to the boundary of $\mathbb{R}^4_+$. To identify where on the boundary the trajectories converge to, it suffices to examine the per-capita growth rate of prey haplotype $1$ at the equilibrium $(0,\widehat n_2^1,\widehat P_2^1)$ and the per-capita growth rate of the prey haplotype $2$ at the equilibrium $(\widehat n_1^1,0,\widehat P_1^1)$. At the first equilibria, we have 

\[
b(1-\widehat n_2^1/K)-d=a_2^1 \widehat P_2^1.
\]

\noindent As $a_2^1>a_1^1$, the per-capita growth rate of prey haplotype $1$ is positive: 

\[
b(1-\widehat n_2^1/K)-d-a_1^1 \widehat P_2^1=(a_2^1-a_1^1) \widehat P_2^1>0.
\]

\noindent Similarly, the per-capita growth rate of prey haplotype $2$ at the other equilibrium is negative:

\[
(a_1^1-a_2^1) \widehat P_1^1<0.
\]

\noindent Hence, the equilibrium $(0,\widehat n_2^1,\widehat P_2^1)$ is unstable while the equilibrium $(\widehat n_1^1,0,\widehat P_1^1)$ is stable. It follows that all trajectories with strictly positive initial conditions converge to this latter equilibrium i.e. haplotype $1$ excludes haplotype $2$ in the presence of predator $1$. A similar argument shows that in the presence of only predator $2$ all solutions with strictly positive initial conditions converge to $(0,\widehat n_2^2,\widehat P_2^2)$ i.e. haplotype $2$ excludes haplotype $1$ in the presence of predator $2$. 

As discussed in the main text, coexistence is only possible if $\widehat n_1^1> \widehat n_1^2$ and $\widehat n_2^2> \widehat n_2^1$. Hence, we assume these inequalities hold. Therefore,  there is a heteroclinic cycle connecting the four boundary equilibria $(\widehat n_1^1,0,\widehat P_1^1,0)$, $(\widehat n_1^2,0,0,\widehat P_1^2)$, $(0,\widehat n_2^2,0,\widehat P_2^2)$, and $(\widehat n_1^2,0,0,\widehat P_1^2)$.  \citep[Theorem 13.6.1, Exercise 13.6.3]{hofbauer-sigmund-98} implies that the heteroclinic cycle is repelling if one can find positive weights $v_1,v_2,w_1,w_2$ such that the function $L(n_1,n_2,P_1,P_2)=n_1^{v_1}n_2^{v_2}P_1^{w_1}P_2^{w_2}$ is an average Lyapunov function i.e. the weighted average of the per-capita growth rates:

\[
\sum_{i=1}^2 v_i \frac{1}{n_i}\frac{dn_i}{dt}+ w_i \frac{1}{P_i}\frac{dP_i}{dt}
\]

\noindent is positive when evaluated at all four of the equilibria along the heteroclinic cycle. Solving the four linear inequalities shows that $v_i>0,w_i>0$ satisfying these conditions exist if and only if the product of the positive per-capita growth rates at these equilibria is greater than the product of the negative per-capita growth rates. Using a similar argument in backwards time (see, e.g., \citep[Lemma 1]{nonlinearity-04}) implies that the heteroclinic cycle is attracting if the product of the positive per-capita growth rates is less than the product of the negative per-capita growth rates. 

The product of the positive per-capita growth rates is   

\[
(a_2^1-a_1^1) \widehat P_2^1 \times (a_1^2-a_2^2) \widehat P_1^2 \times \delta_2 \left(\frac{\widehat n_1^1}{\widehat n_1^2}-1\right)\times \delta_1 \left(\frac{\widehat n_2^2}{\widehat n_2^1}-1\right)
\]

\noindent while the product of the absolute value of the negative per-capita growth rates is 

\[
(a_2^1-a_1^1) \widehat P_1^1
\times (a_1^2-a_2^2) \widehat P_2^2 \times \delta_1 \left(\frac{\widehat n_1^2}{\widehat n_1^1}-1\right)\times \delta_2 \left(\frac{\widehat n_2^1}{\widehat n_2^2}-1\right)
\]

\noindent The first product is greater than the second product if and only if 

\[
 \widehat P_2^1  \widehat P_1^2 \left(\frac{\widehat n_1^1}{\widehat n_1^2}-1\right) \left(\frac{\widehat n_2^2}{\widehat n_2^1}-1\right)
> \widehat P_1^1
 \widehat P_2^2 \left(\frac{\widehat n_1^2}{\widehat n_1^1}-1\right) \left(\frac{\widehat n_2^1}{\widehat n_2^2}-1\right)
\]

\noindent Multiplying both sides by $\widehat n_1^1 \widehat n_2^2 \widehat n_1^2 \widehat n_2^1$ and simplifying yields 

\[
\widehat n_1^1 \widehat n_2^2 \widehat P_2^1  \widehat P_1^2 
>\widehat n_1^2 \widehat n_2^1 \widehat P_1^1
 \widehat P_2^2. 
\]

\noindent Using the explicit expressions for the prey equilibria, we get

\[
\frac{\delta_1}{c_1^1 a_1^1} \frac{\delta_2}{c_2^2 a_2^2} \widehat P_2^1  \widehat P_1^2 
>\frac{\delta_2}{c_1^2 a_1^2} \frac{\delta_1}{c_2^1 a_2^1}  \widehat P_1^1
 \widehat P_2^2. 
\]

\noindent which simplifies to 

\[
\frac{a_1^2 \widehat P_1^2 a_2^1 \widehat P_2^1 }{a_1^1 \widehat P_1^1 a_2^2 \widehat P_2^2} >\frac{c_1^1c_2^2}{c_1^2 c_2^1} =\exp(-\alpha_1-\alpha_2).
\]

\noindent where $a_i^j \widehat P_i^j = b(1-\widehat n_i^j/K)-d$. In the limit of $K\to\infty$,  $\widehat P_i^j= r/a_i^j$ where $r=b-d$. Thus, in this limit, the inequality simplifies to 

\[
\alpha_1+\alpha_2>0.
\]

\begin{suppfigure}[t!]
\begin{center}
\includegraphics[width=0.9\textwidth]{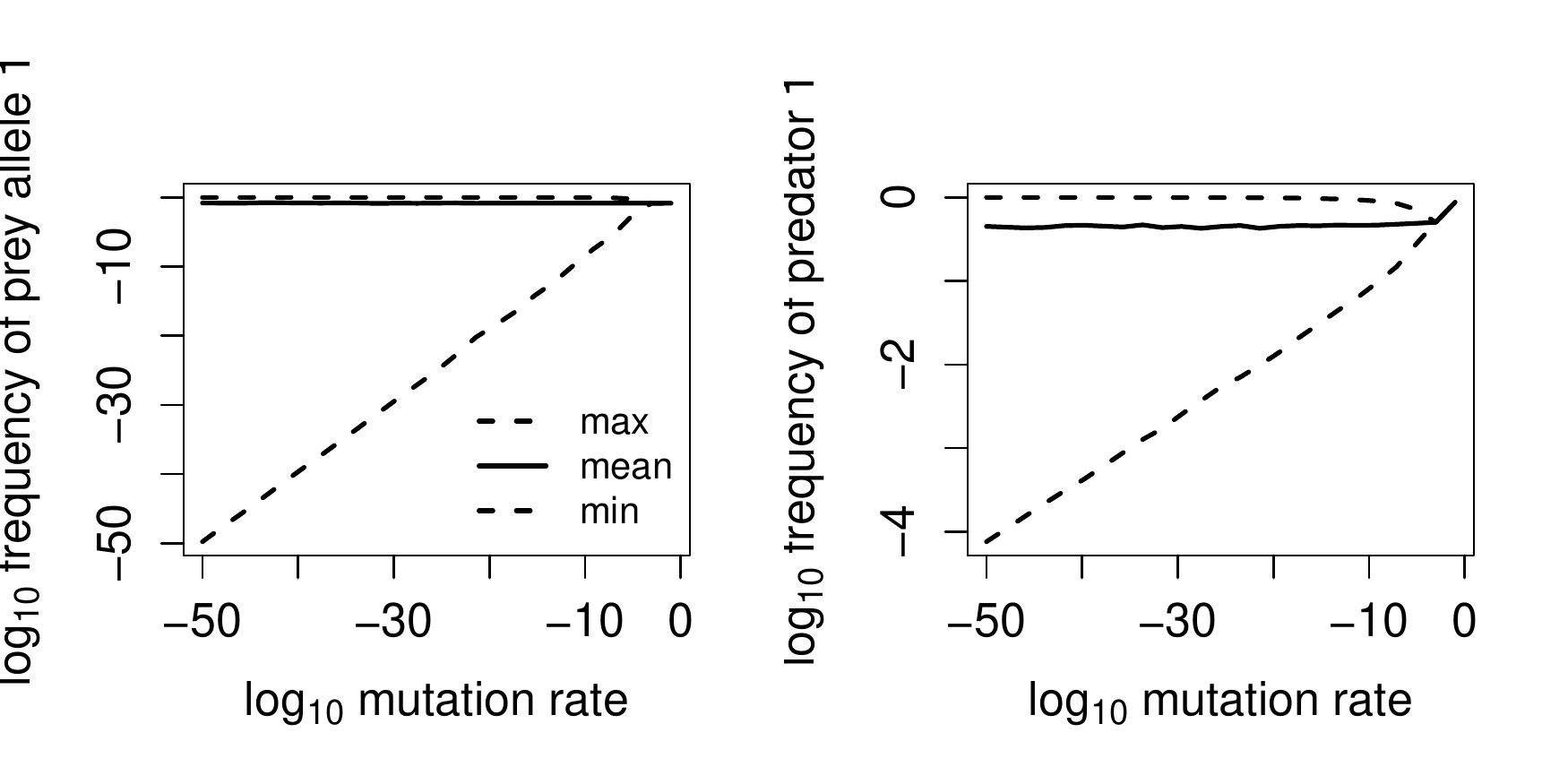}
\caption{Mutation limited coexistence for the case for the haploid model with an attracting heteroclinic cycle. Parameter values: $b=0.5$, $K=500$, $d=0.05$, $\delta_1=0.1$, $\delta_2=0.15$, $a_{1}^1=a_{2}^2=0.1$, $a_{1}^2=a_{2}^1=0.2$, $c_{1}^1=c_{2}^2=0.2$, $c_1^2=c_2^1=0.18$, and the mutation rate $\mu$ varies as shown.}\label{fig:haploid-mutation}
\end{center}
\end{suppfigure}

\paragraph{Including mutation.} If $\mu$ is the probability that one allele mutates to the other, then the haploid dynamics with mutation are given by 

\begin{equation}
\begin{aligned}
\frac{dn_{1}}{dt}&= b(n_1(1-\mu)+n_2\mu)(1-N/K)-dn_{1}-n_1(a_{1}^1 P_1+ a_{1}^2 P_2)\\
\frac{dn_{2}}{dt}&=b(n_1\mu_+n_2(1-\mu))(1-N/K)-dn_{2}- n_2(a_{2}^1 P_1+a_{2}^2 P_2)\\
\frac{dP_1}{dt}&=P_1(c_{1}^1 a_{1}^1 n_{1}+c_{2}^1 a_{2}^1 n_{2}-\delta_1)\\
\frac{dP_2}{dt}&=P_2(c_{1}^2 a_{1}^2 n_{1}+c_{2}^2 a_{2}^2 n_{2}-\delta_2)\\
\end{aligned}
\end{equation}

\noindent Of the different eco-evolutionary assembly scenarios without mutation, only the intransitive case allows for the possibility of coexistence, in the sense of permanence, at small mutation rates. Let us assume the parameter values are consistent with the assembly scenario. Turning on mutations disrupts this intransitivity as all prey genotypes are always present and there are only the subsystems consisting of one predator species and all prey genotypes. Provided the mutation rate is sufficiently low ($\mu \approx 0$), the subsystem with only predator $1$ (respectively $2$) has a globally stable, feasible equilibrium at which $(n_1, n_2, P_1)\approx (\widehat n_1^1, 0, \widehat P_1^1)$ (respectively,  $(n_1, n_2, P_2)\approx ( 0,\widehat n_2^2, \widehat P_2^2)$)  up to order $\rm{O}(\mu)$. The per-capita growth rate of predator $2$ at this equilibrium is given by (up to order $\rm{O}(\mu)$)

\[
c_1^2 a_1^2 \widehat n_1^1 -\delta_2 = \delta_2\left(\frac{\widehat n_1^1}{\widehat n_1^2}-1\right) 
\]

\noindent which is positive by the intransitivity conditions. Similarly, the per-capita growth rate of predator $1$ is positive at the equilibrium only supporting predator $2$. Thus, applying classical results from permanence theory e.g. \citep{garay-89}, we have verified the claim in the main text that the species always coexist in the sense of permanence provided there are mutations and the intransitivity condition is satisfied. However, if the coexistence condition in the absence of mutations is not satisfied, the heteroclinic cycle corresponding to the intransitivity is an attractor for the mutation free system. Upper semi-continuity of attractors~\citep{conley-78} implies there is an attractor arbitrarily close to the boundary of $\mathbb{R}_+^4$ provided $\mu$ is sufficiently small. Fig.~\ref{fig:haploid-mutation} shows how the minimal frequencies of the predator species or the prey genotypes depend on the mutation rate $\mu$ when the heteroclinic cycle is attracting for the model without mutations. Consistent with the analytic predictions, the minimal log-frequencies are on the order of $\log \mu$.

\subsection*{The diploid case}
As with the haploid case, we begin with the subsystem consisting of all three prey genotypes and a single predator, say predator $1$, and show that allele $A_1$ fixates. Define $n_i=2n_{ii}+n_{12}$ be the density of alleles $A_i$ in the population. We show that the function $V(n_{11},n_{12},n_{22},P_1)=\log n_1 - \log n_2$ increases along solutions of this subsystem whenever $n_2>0$ and $P_1>0$ from which it follows that $(\widehat n_{11}^1,0,0,\widehat P_{11}^1)$ is globally stable in this subsystem. A similar argument implies that $(0, 0, \widehat n_{22}^2, \widehat P_{22}^2)$ is globally stable in the $n_{11}$, $n_{12}$, $n_{22}$, and $P_2$ subsystem. 

To prove our assertion, assume that both alleles are present (i.e. $n_1>0$, $n_2>0$), $P_1>0$, and $P_2=0$. Then 

\[
\begin{aligned}
\frac{dV}{dt}=&\frac{1}{n_1}\frac{dn_1}{dt}-\frac{1}{n_2}\frac{dn_2}{dt}\\
=&\frac{1}{n_1}\left(bN(2(x_{11})^2+2x_{11}x_{12}+(x_{12})^2/2+x_{11}x_{12}+2x_{11}x_{22}+x_{22}x_{12}+(x_{12})^2/2)(1-N/K)\right)\\
&-\frac{1}{n_1}\left(dn_1+ 2a_{11}^1 n_{11}P_1+a_{12}^1n_{12}P_1\right)\\
&-\frac{1}{n_2}\left(bN(2(x_{22})^2+2x_{22}x_{12}+(x_{12})^2/2+x_{22}x_{12}+2x_{11}x_{22}+x_{11}x_{12}+(x_{12})^2/2)(1-N/K)\right)\\
&+\frac{1}{n_2}\left(dn_2+ 2a_{22}^1 n_{22}P_1+a_{12}^1n_{12}P_1\right)\\
&=\frac{1}{n_1}\left(bn_1(1-N/K)-dn_1- 2a_{11}^1 n_{11}P_1-a_{12}^1n_{12}P_1\right)\\
&-\frac{1}{n_2}\left(bn_2(1-N/K)-dn_2- 2a_{22}^1 n_{22}P_1-a_{12}^1n_{12}P_1\right)\\
&=P_1 \left(\frac{2a_{22}^1 n_{22}+a_{12}^1n_{12}}{n_2}-\frac{2a_{11}^1 n_{11}+a_{12}^1n_{12}}{n_1}\right)
\end{aligned}
\]

\noindent which is strictly positive due to our assumption that $a_{11}^1<a_{12}^1<a_{22}^1$. Hence,  $V$ increases along solutions with initial conditions satisfying $n_1>0,n_2>0, P_1>0$ and $P_2=0$. From this it follows that $(\widehat n_{11}^1,0,0,\widehat P_{11}^1)$ is globally stable in this subsystem. We also note that the equilibrium $E_{11}^1=(\widehat n_{11}^1,0,0,\widehat P_{11}^1)$ is linearly stable and the equilibrium $E_{22}^1=(0,0,\widehat n_{22}^1,\widehat P_{22}^1)$ is linearly unstable for the prey-predator $1$ subsystem. Indeed, the per-capita growth rate of allele $2$ at $E_{11}^1$ is 

\[
\begin{aligned}
\frac{1}{n_2}\frac{dn_2}{dt}\Big|_{E_{11}^1}&=b(1-\widehat n_{11}^1/K)-d-a_{12}^1\widehat P_{11}^1\\
&=\left(a_{11}^1-a_{12}^1\right)\widehat P_{11}^1<0
\end{aligned}
\]

\noindent and, consequently, this equilibrium is stable in the prey-predator $1$ subsystem. Similarly, the per-capita growth rate of allele $1$ at $E_{22}^1$ equals

\[
\left(a_{22}^1-a_{12}^1\right)\widehat P_{22}^1>0
\]

\noindent and, consequently, this equilibrium is unstable in the prey-predator $1$ subsystem. 

As in the haploid case, we have a heteroclinic cycle between $4$ equilibria on the boundary: $(n_{11},n_{12},n_{22},P_1,P_2)=(\widehat n_{11}^1,0,0,\widehat P_{11}^1,0)$, $(\widehat n_{11}^2,0,0,0,\widehat P_{11}^2)$, $(0,0,\widehat n_{22}^2,0,\widehat P_{22}^2)$, and $(0,0,\widehat n_{22}^1,\widehat P_{22}^1,0)$. To determine when this heteroclinic cycle is repelling or attracting, we use the function $L=n_1^{x_1} n_2^{x_2} P_1^{y_1} P_2^{y_2}$ where $n_i$ is the density of allele $i$ in the prey population. $L$ is an average Lyapunov function (see \citet[Section 12.2]{hofbauer-sigmund-98}) if we can find $x_i>0$ and $y_i>0$ such that 

\[
\sum_i x_i \frac{1}{n_i}\frac{dn_i}{dt}+y_i \frac{1}{P_i}\frac{dP_i}{dt}>0
\]

\noindent at all the equilibria of the heteroclinic cycle. A standard calculation involving these linear inequalities implies that there is a solution if and only if the product of the positive per-capita growth rates at these equilibria is greater than the product of the absolute value of the negative per-capita growth rates at these equilibria. When this occurs, \citet[Theorem 12.2.1]{hofbauer-sigmund-98} implies that the heteroclinic cycle is repelling. 

The product of the positive per-capita growth rates is 

\[
(a_{22}^1-a_{12}^1) \widehat P_{22}^1 \times (a_{11}^2-a_{12}^2) \widehat P_{11}^2 \times \delta_2 \left(\frac{\widehat n_{11}^1}{\widehat n_{11}^2}-1\right)\times \delta_1 \left(\frac{\widehat n_{22}^2}{\widehat n_{22}^1}-1\right)
\]

\noindent while the product of the absolute value of the negative per-capita growth rates is  

\[
(a_{12}^1-a_{11}^1) \widehat P_{11}^1
\times (a_{12}^2-a_{22}^2) \widehat P_{22}^2 \times \delta_1 \left(\frac{\widehat n_{11}^2}{\widehat n_{11}^1}-1\right)\times \delta_2 \left(\frac{\widehat n_{22}^1}{\widehat n_{22}^2}-1\right).
\]

\noindent Multiplying both sides of the inequality by $\widehat n_{11}^1\widehat n_{11}^2 \widehat n_{22}^1 \widehat n_{22}^2$ and canceling like terms, we get 

\[
(a_{22}^1-a_{12}^1)(a_{11}^2-a_{12}^2) \widehat P_{22}^1  \widehat P_{11}^2  \widehat n_{11}^1 \widehat n_{22}^2 >
(a_{12}^1-a_{11}^1) (a_{12}^2-a_{22}^2)\widehat P_{11}^1 \widehat P_{22}^2 \widehat n_{11}^2 \widehat n_{22}^1.
\]

\noindent Using the definition of $\widehat n_{ii}^j=\frac{\delta_j}{a_{ii}^jc_{ii}^j}$ and simplifying yields our coexistence condition:

\[
\exp(\beta_1+\beta_2)
 \frac{ a_{22}^1 \widehat P_{22}^1 a_{11}^2 \widehat P_{11}^2  }{a_{11}^1\widehat P_{11}^1 a_{22}^2\widehat P_{22}^2} >\exp(-\alpha_1-\alpha_2).
\]

\noindent In limit of $K=\infty$, $\widehat P_{ii}^\ell$ equals $r/a_{ii}^\ell$ where $r=b-d$, and  the inequality simplifies to 

\[
\beta_1+\beta_2+\alpha_1+\alpha_2>0.
\]

\paragraph{Including mutation.}  Let $\mu$ be the probability the one allele mutates to the other, and $\nu=1-\mu$. The diploid model with mutation is given by 

\begin{equation}
\begin{aligned}
\frac{dn_{11}}{dt}&=bN(\underbrace{(x_{11})^2\nu^2+x_{11}x_{12}\nu +2 x_{11}x_{22}\mu\nu+(x_{22})^2\mu^2+(\nu^2/4+\mu\nu/2+\mu^2) (x_{12})^2}_{\Psi_1})(1-N/K)\\
&-dn_{11}- a_{11}^1 n_{11}P_1-a_{11}^2 n_{11}P_2\\
\frac{dn_{22}}{dt}&=bN(\underbrace{(x_{22})^2\nu^2+x_{22}x_{12}\nu +2 x_{11}x_{22}\mu\nu+(x_{11})^2\mu^2+(\nu^2/4+\mu\nu/2+\mu^2)(x_{12})^2}_{\Psi_2})(1-N/K)\\
&-dn_{22}-a_{22}^1 n_{22}P_1-a_{22}^2 n_{22}P_2\\
\frac{dn_{12}}{dt}&=bN(1-\Psi_1-\Psi_2)(1-N/K)-dn_{12}-a_{12}^1 n_{12}P_1-a_{12}^2 n_{12}P_2\\
\frac{dP_1}{dt}&=P_1(c_{11}^1 a_{11}^1 n_{11}+c_{12}^1 a_{12}^1 n_{12}+c_{22}^1 a_{22}^1 n_{22}-\delta_1)\\
\frac{dP_2}{dt}&=P_2(c_{11}^2 a_{11}^2 n_{11}+c_{12}^2 a_{12}^2 n_{12}+c_{22}^2 a_{22}^1 n_{22}-\delta_2)\\
\end{aligned}
\end{equation}

\noindent As in the haploid case, of the different eco-evolutionary assembly scenarios without mutation, only the intransitive case allows for  permanence. Let us assume the parameter values are consistent with the assembly scenario. Turning on mutations disrupts this intransitivity as all prey genotypes are present and there are only the subsystems consisting of one predator species and all prey genotypes. Provided the mutation rate is sufficiently low ($\mu \approx 0$), the subsystem with only predator $1$ (respectively $2$) has a globally stable, feasible equilibrium at which $(n_{11},n_{12}, n_{22}, P_1)\approx (\widehat n_{11}^1, 0,0, \widehat P_{11}^1)$ (respectively,  $(n_{11}, n_{12},n_{22}, P_2)\approx ( 0,0,\widehat n_{22}^2, \widehat P_{22}^2)$)  up to order $O(\mu)$. The per-capita growth rate of predator $2$ at this equilibrium is given by (up to order $O(\mu)$)
\[
c_{11}^2 a_{11}^2 \widehat n_{11}^1 -\delta_2 = \delta_2\left(\frac{\widehat n_{11}^1}{\widehat n_{11}^2}-1\right)>0 
\]
by the intransitivity conditions. Similarly, the per-capita growth rate of predator $1$ is positive at the equilibrium only supporting predator $2$.  Thus, applying classical results from permanence theory e.g. \citep{garay-89}, we have verified the claim in the main text that the species always coexist in the sense of permanence provided there are mutations and the intransitivity condition is satisfied. However, if the coexistence condition in the absence of mutations is not satisfied, the heteroclinic cycle for the mutation free model is an attractor. Upper semi-continuity of attractors implies that the system with sufficiently small mutation rates has an attractor arbitrarily close to the boundary of $\mathbb{R}^5_+$.

\section{Analysis of the apparent competition models}\label{AppendixB}
For the apparent competition module, the haploid model is given by 

\begin{equation}
\begin{aligned}
\frac{dN_{1}}{dt}&=N_1(r_1(1-N_1/K_1)- a_{1}^1 p_1- a_{1}^2 p_2)\\
\frac{dN_{2}}{dt}&=N_2(r_2(1-N_2/K_2)- a_{2}^1 p_1- a_{2}^2 p_2)\\
\frac{dp_1}{dt}&=p_1(c_{1}^1 a_{1}^1 N_{1}+c_{2}^1 a_{2}^1 N_{2}-\delta)\\
\frac{dp_2}{dt}&=p_2(c_{1}^2 a_{1}^2 N_{1}+c_{2}^2 a_{2}^2 N_{2}-\delta)\\
\end{aligned}
\end{equation}

\noindent while the diploid model is 

\begin{equation}
\begin{aligned}
\frac{dN_{1}}{dt}&=N_1(r_1(1-N_1/K_1)- a_{1}^{11} p_{11}- a_{1}^{22} p_{22}-a_1^{12}p_{12})\\
\frac{dN_{2}}{dt}&=N_2(r_2(1-N_2/K_2)- a_{2}^{11} p_{11}- a_{2}^{22} p_{22}-a_2^{12}p_{12})\\
\frac{dp_{11}}{dt}&=(c_{1}^{11} a_{1}^{11}N_1+c_2^{11}a_2^{11} N_2)p_{11}(x_{11}+x_{12}/2)+(c_{1}^{12} a_{1}^{12}N_1+c_2^{12}a_2^{12} N_2)p_{12}(x_{11}/2+x_{12}/4)-\delta p_{11}\\
\frac{dp_{22}}{dt}&=(c_{1}^{22} a_{1}^{22}N_1+c_2^{22}a_2^{22} N_2)p_{22}(x_{22}+x_{12}/2)+(c_{1}^{12} a_{1}^{12}N_1+c_2^{12}a_2^{12} N_2)p_{12}(x_{22}/2+x_{12}/4)-\delta p_{22}\\
\frac{dp_{12}}{dt}&=(c_{1}^{11} a_{1}^{11}N_1+c_2^{11}a_2^{11} N_2)p_{11}(x_{22}+x_{12}/2)+(c_{1}^{12} a_{1}^{12}N_1+c_2^{12}a_2^{12} N_2)p_{12}(x_{11}/2+x_{22}/2+x_{12}/2)\\
&+(c_{1}^{22} a_{1}^{22}N_1+c_2^{22}a_2^{22} N_2)p_{22}(x_{11}+x_{12}/2)-\delta p_{12}
\end{aligned}
\end{equation}

\noindent Our analysis will focus on the diploid model as the conditions for coexistence in the haploid model agree with the diploid model with additive genetics (i.e. $a_i^{12}=(a_i^{11})+a_i^{22})/2$ and $c_i^{12}a_i^{12}=(c_i^{11}a_i^{11}+c_i^{22}a_i^{22})/2$).

We begin with the subsystem consisting of all three predator genotypes and a single prey species, say prey $1$. We will show that the homozygote $A_1A_1$  excludes the other predator genotypes. Define $p_i=2p_{ii}+p_{12}$ be the density of alleles $A_i$ in the population. We will show that the function $V(N_1,p_{11},p_{12},p_{22})=\log p_1 - \log p_2$ increases along solutions of this subsystem whenever $N_1>0$ and $p_1>0$ from which it follows that $(\widehat N_1^{11},\widehat p_1^{11},0,0)$ is globally stable in this subsystem where $\widehat N_1^{11}=\delta/(c_1^{11}a_1^{11})$ and $\widehat p_1^{11}=(b_1(1-\widehat N_1^{11}/K_1)-d_1)/a_1^{11}$. A similar argument implies that the $A_2$ alleles sweep to fixation whenever only prey species $2$ is present. 

To prove our assertion, assume that $p_1>0$, $p_2>0$ (i.e. both alleles are present), $N_1>0$, and $N_2=0$. Define $P=p_{11}+p_{12}+p_{22}$ as the total predator density, $x_{ij}=p_{ij}/P$ as the frequency of genotype $A_iA_j$, and $x_i=p_i/(2P)$ as the frequency of allele $A_i$. Then 

\[
\begin{aligned}
\frac{dV}{dt}=&\frac{1}{p_1}\frac{dp_1}{dt}-\frac{1}{p_2}\frac{dp_2}{dt}\\
=&\frac{1}{p_1}\left(c_1^{11}a_1^{11}p_{11}(1+x_1)+c_1^{12}a_1^{12}p_{12}(1/2+x_1)+c_1^{22}a_1^{22}p_{22}x_1\right)N_1\\
-&\frac{1}{p_2}\left(c_1^{22}a_1^{22}p_{22}(1+x_2)+c_1^{12}a_1^{12}p_{12}(1/2+x_2)+c_1^{11}a_1^{11}p_{11}x_2\right)N_1\\
=&\left(c_1^{11}a_1^{11}p_{11}\left(\frac{1}{p_1}+\frac{1}{2P}\right)+c_1^{12}a_1^{12}p_{12}\left(\frac{1}{2p_1}+\frac{1}{2P}\right)+c_1^{22}a_1^{22}p_{22}\frac{1}{2P}\right)N_1\\
-&\left(c_1^{22}a_1^{22}p_{22}\left(\frac{1}{p_2}+\frac{1}{2P}\right)+c_1^{12}a_1^{12}p_{12}\left(\frac{1}{2p_2}+\frac{1}{2P}\right)+c_1^{11}a_1^{11}p_{11}\frac{1}{2P}\right)N_1\\
=&\left(c_1^{11}a_1^{11}\frac{p_{11}}{p_1}+c_1^{12}a_1^{12}\frac{p_{12}}{2p_1}\right)N_1-\left(c_1^{22}a_1^{22}\frac{p_{22}}{p_2}+c_1^{12}a_1^{12}\frac{p_{12}}{2p_2}\right)N_1
\end{aligned}
\]

\noindent which is strictly positive as the first term is a strict convex combination of $c_1^{11}a_1^{11}$ and $c_1^{12}a_{1}^{12}$, the second term is a strict convex combination of $c_1^{22}a_1^{22}$ and $c_1^{12}a_1^{12}$, and $c_1^{11}a_1^{11}>c_1^{12}a_1^{12}>c_1^{22}a_1^{22}$. Hence, allele $1$ sweeps to fixation whenever only prey $1$ is present. Similarly, allele $2$ sweeps to fixation whenever only prey $2$ is present.

For the subsystem consisting of one predator homozygous genotype and two prey, the ecological theory of apparent competition applies~\citep{holt-77}. When the $K_i$ are sufficiently large, the prey species which supports the higher equilibrium predator density excludes the other prey species. A formal proof of this statement follows from Theorem 6 of \citet{takeuchi-adachi-83}. For the remainder of this Appendix, we assume that the $K_i$ are sufficiently large so that one prey is excluded whenever one of the predator alleles is present. These observations about the subsystem dynamics lead to three types of eco-evolutionary assembly diagrams shown in Figure~\ref{fig:assemble-apparent}.

Only in the case of an intransitivity is coexistence, in the sense of permanence, possible. We examine this condition in the limit of high $K_1,K_2$ values. The stability of the heteroclinic cycle is determined by examining the products of the positive and negative per-capita growth rates at the equilibria. The per-capita growth rate of prey $j$ at the equilibrium determined by prey $i\neq j$ and predator allele $A_\ell$ is given by 

\[
r_j - a_{j}^{\ell \ell }\widehat p_i^{\ell\ell} =  a_j^{\ell\ell}(\widehat p_j^{\ell\ell} -\widehat p_i^{\ell\ell})
\]

\noindent which is positive if $i=\ell$ and negative otherwise. The per-capita growth rate of predator allele $j$ at the equilibrium determined by prey $i$ and predator allele $\ell \neq j$ is 

\[
c_i^{12}a_i^{12}\widehat N_i^{\ell\ell}-\delta
= \frac{\delta}{c_i^{\ell\ell}a_i^{\ell\ell}}\left(
c_i^{12}a_i^{12}-c_i^{\ell\ell}a_i^{\ell\ell}
\right)
\]

which is positive if $i\neq \ell$ and negative otherwise. Therefore, the product of the positive per-capita growth rates is 

\[
a_2^{11}(\widehat p_2^{11} -\widehat p_1^{11})a_1^{22}(\widehat p_1^{22} -\widehat p_2^{22})\frac{\delta}{c_1^{22}a_1^{22}}\left(
c_1^{12}a_1^{12}-c_1^{22}a_1^{22}\right)\frac{\delta}{c_2^{11}a_2^{11}}\left(
c_2^{12}a_2^{12}-c_2^{11}a_2^{11}
\right)
\]

\noindent and the product of the negative per-capita growth rate is 

\[
a_2^{22}(\widehat p_2^{22} -\widehat p_1^{22})a_1^{11}(\widehat p_1^{11} -\widehat p_2^{11})\frac{\delta}{c_1^{11}a_1^{11}}\left(
c_1^{12}a_1^{12}-c_1^{11}a_1^{11}
\right)\frac{\delta}{c_2^{22}a_2^{22}}\left(
c_2^{12}a_2^{12}-c_2^{22}a_2^{22}
\right)
\]

\noindent The product of the positive per-capita growth rates is greater than the product of the negative per-capita growth rates if and only if 

\[
\frac{1}{c_1^{22}}\left(
c_1^{12}a_1^{12}-c_1^{22}a_1^{22}\right)\frac{1}{c_2^{11}}\left(
c_2^{12}a_2^{12}-c_2^{11}a_2^{11}
\right)>
\frac{1}{c_1^{11}}\left(
c_1^{12}a_1^{12}-c_1^{11}a_1^{11}
\right)\frac{1}{c_2^{22}}\left(
c_2^{12}a_2^{12}-c_2^{22}a_2^{22}
\right)
\]

\noindent Equivalently,

\[
\exp(\beta_1+\beta_2)=
\frac{
|c_1^{12}a_1^{12}-c_1^{22}a_1^{22}|}{|c_1^{12}a_1^{12}-c_1^{11}a_1^{11}|}\frac{|c_2^{12}a_2^{12}-c_2^{11}a_2^{11}|}{|c_2^{12}a_2^{12}-c_2^{22}a_2^{22}|}
>
\frac{c_1^{22}}{c_1^{11}}\frac{c_2^{11}}{c_2^{22}}=\exp(-\alpha_1-\alpha_2)
\]

\noindent as claimed in the main text.
\begin{suppfigure}[t!]
\begin{center}
\includegraphics[width=0.9\textwidth]{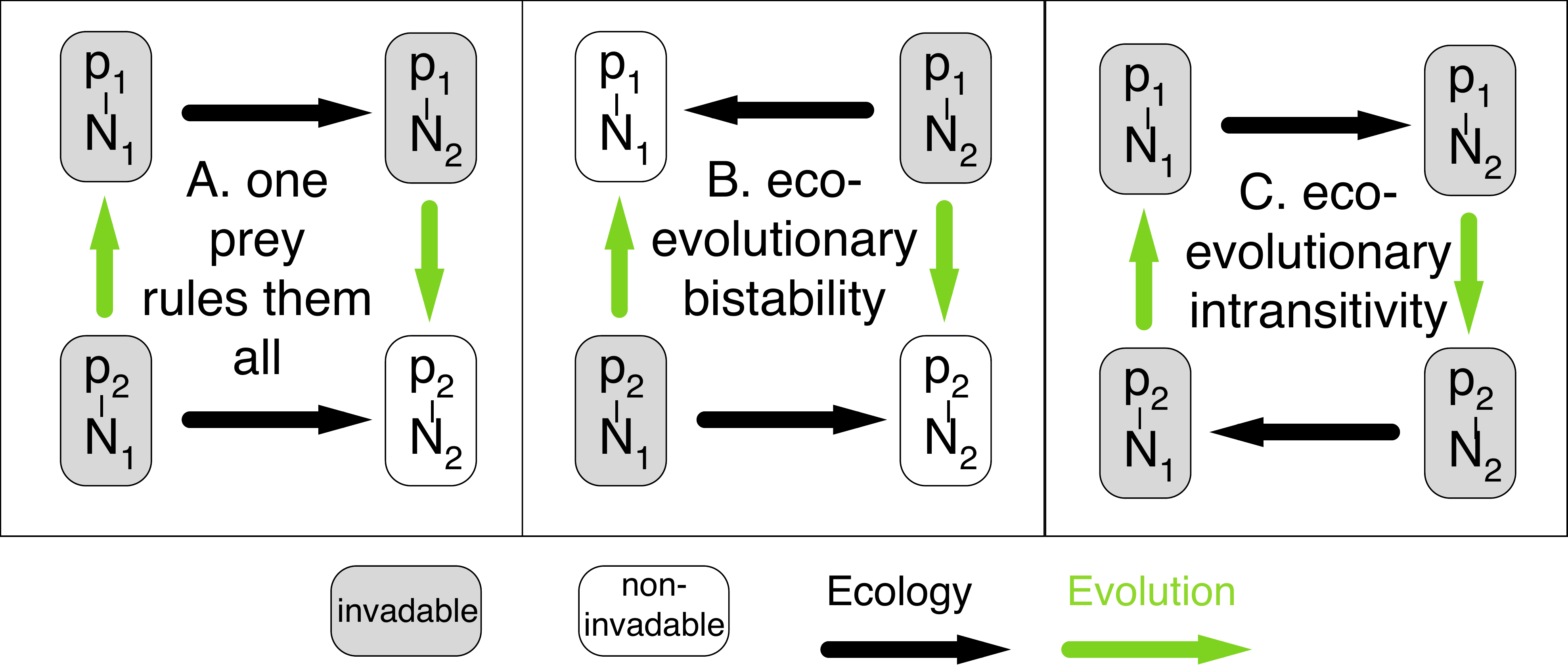}
\end{center}
\caption{The eco-evolutionary assembly diagrams for the apparent competition model. The node of each diagram corresponds to an equilibrium of a subcommunity of species and genotypes. Black arrows correspond to transitions between subcommunities due predator invasions. Green arrows correspond to transitions due to invasions of prey alleles. Stable (i.e. non-invadible) communities are shown as white boxes, others are gray. In A, one predator is able to suppress both homozygous prey genotypes to a lower equilibrium density than the other predator species. In B, each homozygous prey genotype is suppressed to the lower equilibrium density by the predator to which it is least defended. In C, each homozygous prey genotype is suppressed to the lower equilibrium density by the predator to which it is most defended.}\label{fig:assemble-apparent}
\end{suppfigure}

\end{document}